\documentclass[%
 aip,
 amsmath,amssymb,
preprint,%
]{revtex4-1}

\usepackage{graphicx}% Include figure files
\usepackage{dcolumn}% Align table columns on decimal point
\usepackage{bm}% bold math
\usepackage[utf8]{inputenc}
\usepackage[T1]{fontenc}
\usepackage{mathptmx}
\usepackage{etoolbox}

\makeatletter
\def\@email#1#2{%
 \endgroup
 \patchcmd{\titleblock@produce}
  {\frontmatter@RRAPformat}
  {\frontmatter@RRAPformat{\produce@RRAP{*#1\href{mailto:#2}{#2}}}\frontmatter@RRAPformat}
  {}{}
}%
\makeatother
\newtheorem{thm}{Theorem}[section]
\newtheorem{rem}{Remark}[section]

\newcommand{\dE}{\mathbb{E}}
\newcommand{\dR}{\mathbb{R}}

\newcommand{\dP}{\mathbb{P}}
\newcommand{\dZ}{\mathbb{Z}}

\newcommand{\cN}{\mathcal{N}}

\newcommand{\cV}{\mathcal{V}}
\newcommand{\rI}{\openone}
\newcommand{\cF}{\mathcal{F}}

\newcommand{\cR}{\mathcal{R}}

\newcommand{\cM}{\mathcal{M}}

\newcommand{\ind}{\mbox{1}\kern-.25em \mbox{I}}
\font\calcal=cmsy10 scaled\magstep1
\def\build#1_#2^#3{\mathrel{\mathop{\kern 0pt#1}\limits_{#2}^{#3}}}
\def\liml{\build{\longrightarrow}_{}^{{\mbox{\calcal L}}}}

\def\videbox{\mathbin{\vbox{\hrule\hbox{\vrule height1.4ex \kern.6em\vrule height1.4ex}\hrule}}}
\def\demend{\hfill $\videbox$\\}

\date{\today}

\makeatletter
\def\@email#1#2{%
 \endgroup
 \patchcmd{\titleblock@produce}
  {\frontmatter@RRAPformat}
  {\frontmatter@RRAPformat{\produce@RRAP{*#1\href{mailto:#2}{#2}}}\frontmatter@RRAPformat}
  {}{}
}%
\makeatother
\begin{document}

\preprint{AIP/123-QED}

\title[Asymptotic analysis of random walks on ice and graphite]
{Asymptotic analysis of random walks on ice and graphite}

\author{Bernard Bercu}

\affiliation{Institut de Math\'ematiques de Bordeaux, Universit\'e de Bordeaux, 
UMR 5251, 351 Cours de la Lib\'eration, 33405 Talence cedex, France.}%

\author{Fabien Montégut}
 \email{fabien.montegut@math.univ-toulouse.fr.}
\affiliation{%
Institut de Math\'ematiques de Toulouse,  Universit\'e de Toulouse,
UMR 5219, 118 Route de Narbonne, 31062 Toulouse cedex, France.
}%

\date{\today}

\begin{abstract}
The purpose of this paper is to investigate the asymptotic behavior of random walks on three-dimensional crystal
structures. We focus our attention on the $1h$ structure of the ice and the $2h$ structure of graphite.
We establish the strong law of large numbers and the asymptotic normality for both random walks on ice and graphite.
All our analysis relies on asymptotic results for multi-dimensional martingales.
\end{abstract}

\keywords{Random walk, Hexagonal lattice, Central limit theorem \\ \textbf{MSC (2010)} Primary: 60G50; Secondary: 60F05; 82C41}

\maketitle

%%%%%%%%%%%%%%%%%%%%%%%%%%%%%%%%%%%%%%%%%%%%%%%%%%%%%%%%%%%%%%%%%%%%%%%%%%%%%%%%%%%%%%%%%%%%%%%%%%%%%%%%%%%%%%%%%%%%%%%%
\section{Introduction}
\label{S-I}
%%%%%%%%%%%%%%%%%%%%%%%%%%%%%%%%%%%%%%%%%%%%%%%%%%%%%%%%%%%%%%%%%%%%%%%%%%%%%%%%%%%%%%%%%%%%%%%%%%%%%%%%%%%%%%%%%%%%%%%%

A wide variety of materials present a repeating symmetrical arrangement of their atoms, molecules or ions,  
known as crystal structures. Those underlying structures determine some physical properties such as the toughness, the porosity or even the
conductivity of the materials. They can be looked further upon by studying the behavior of random walks in the crystal structures, 
see e.g. (\onlinecite{BZ82}) for a study of energy trapping in crystal structure or (\onlinecite{Kovacik96}) for the electrical conductivity of Cu-graphite 
composites. In particular, random walks are widely used to determine the diffusion of vacancies or point defects in crystals \cite{Koiwa}.

Random walks represent a large class of Markov chains and several reference books \cite{Feller,Rudnick} are 
devoted to the study of their properties, such as the probability of returning to their starting point, the shape of typical trajectories or 
their long-time behavior. Polya \cite{Polya} was the first to observe the influence of the dimension of the lattice on their properties, 
as the simple random walk on $\dZ^d$ becomes transient when $d\geq 3$. The model of a non simple random walk on periodic lattice is quite convenient to study the properties of crystalline solids as stated in (\onlinecite{Montroll}).

Cubic crystal structures were previously studied in terms of random walks \cite{Garza} or vacancy diffusions \cite{Bocquet}, 
see also (\onlinecite{deForcrand}) and (\onlinecite{GarzaHexa}) for planar honeycomb lattices.
However, to the best of our knowledge, three-dimensional hexagonal lattices still have to be considered. The goal of this paper is to investigate the asymptotic behavior of random walks in two hexagonal crystal structures in three dimensions, namely the $1h$ structure of the ice and the $2h$ structure of graphite. Both of them can be seen as sheets of infinite hexagonal plane lattices stacked on top 
of each other, where the way the consecutive sheets are stacked drastically changes the properties of the structure.

On the one hand, the properties of ice are theoretical and experimental research subjects since decades, see e.g. the pioneering works (\onlinecite{Bradley57}) or (\onlinecite{DMS64}). On the other hand, graphite composites finds many applications in a wide range of fields, see e.g. (\onlinecite{Inagaki89}) as well as 
the references therein, and its $2h$ structure may be found in other materials \cite{Perevislov19}. In both cases, 
understanding the asymptotic behavior of random walks in such structures is a key step in unveiling some of these materials properties.

Random walks on the two-dimensional hexagonal structure of the graphene has been described many times, especially in (\onlinecite{Crescenzo19}) where 
the authors studied the large deviation properties of the random walk, using a parity argument based on the structure of such lattice. 
Our purpose is to extend several results in (\onlinecite{Crescenzo19}) to the three-dimensional hexagonal structures we are interested in.
In this paper, we assume that the transition probabilities are invariant by translating the unit cell of the crystal.
Our goal is to establish the strong law of large numbers and the asymptotic normality of the random walk in both structures. 

Our strategy is to separate the vertices of the lattice depending on their local geometry. In the simple case of the random walk on ice (RWI), 
there are only two different types of vertices. On the contrary, the random walk on the graphite (RWG) admits four different types 
of vertices and this situation is much more difficult to handle. 

The paper is organized as follows: the definition and description of the random walks and their transition probabilities are given in Section 2. 
Section 3 is devoted to our main results. To be more precise, we establish the strong law of large numbers and the asymptotic normality 
for both RWI and RWG. The results concerning the RWI are proven in Section 4, while their counterparts for the RWG are postponed to Section 5. 
All our analysis relies on asymptotic results for multi-dimensional martingales.
Finally, Section 6 contains concluding remarks and perspectives. 

%%%%%%%%%%%%%%%%%%%%%%%%%%%%%%%%%%%%%%%%%%%%%%%%%%%%%%%%%%%%%%%%%%%%%%%%%%%%%%%%%%%%%%%%%%%%%%%%%%%%%%%%%%%%%%%%%%%%%%%%
\section{Two possible structures}
\label{S-G}
%%%%%%%%%%%%%%%%%%%%%%%%%%%%%%%%%%%%%%%%%%%%%%%%%%%%%%%%%%%%%%%%%%%%%%%%%%%%%%%%%%%%%%%%%%%%%%%%%%%%%%%%%%%%%%%%%%%%%%%%

\begin{figure}[htbp]
    \centering
    \vspace{-1cm}
\includegraphics[scale=0.2]{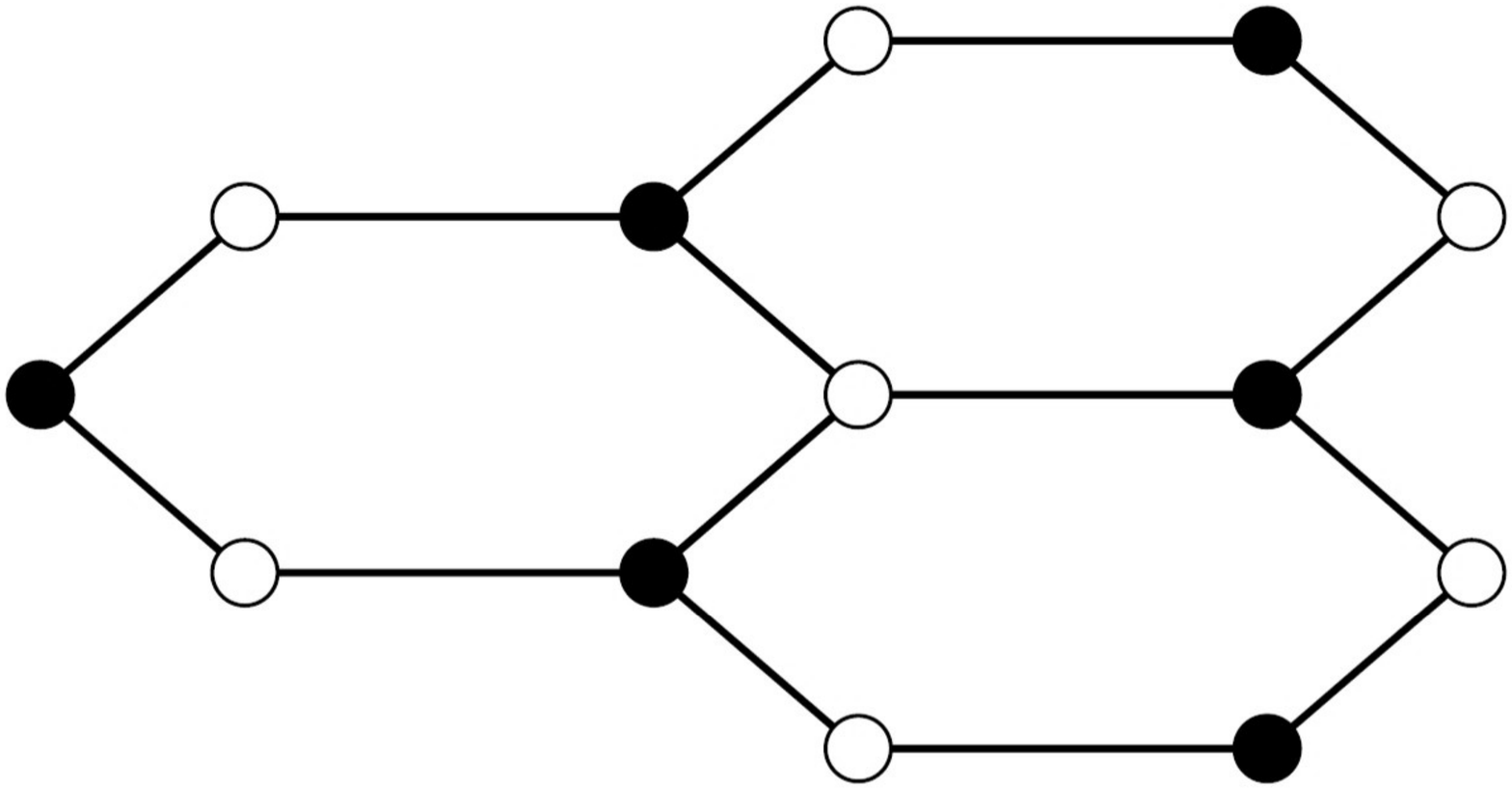}
\vspace{-1cm}
    \caption{Hexagonal structure of the graphene}
    \label{FigGraphene}
\end{figure}

The two-dimensional hexagonal structure of the graphene was previously considered in (\onlinecite{Crescenzo19}) 
where two different kind of vertices $\cV_0$ and $\cV_1$ are represented in Figure \ref{FigGraphene} with white and black circles.

\begin{figure}[htbp]
    \centering
\includegraphics[scale=0.2]{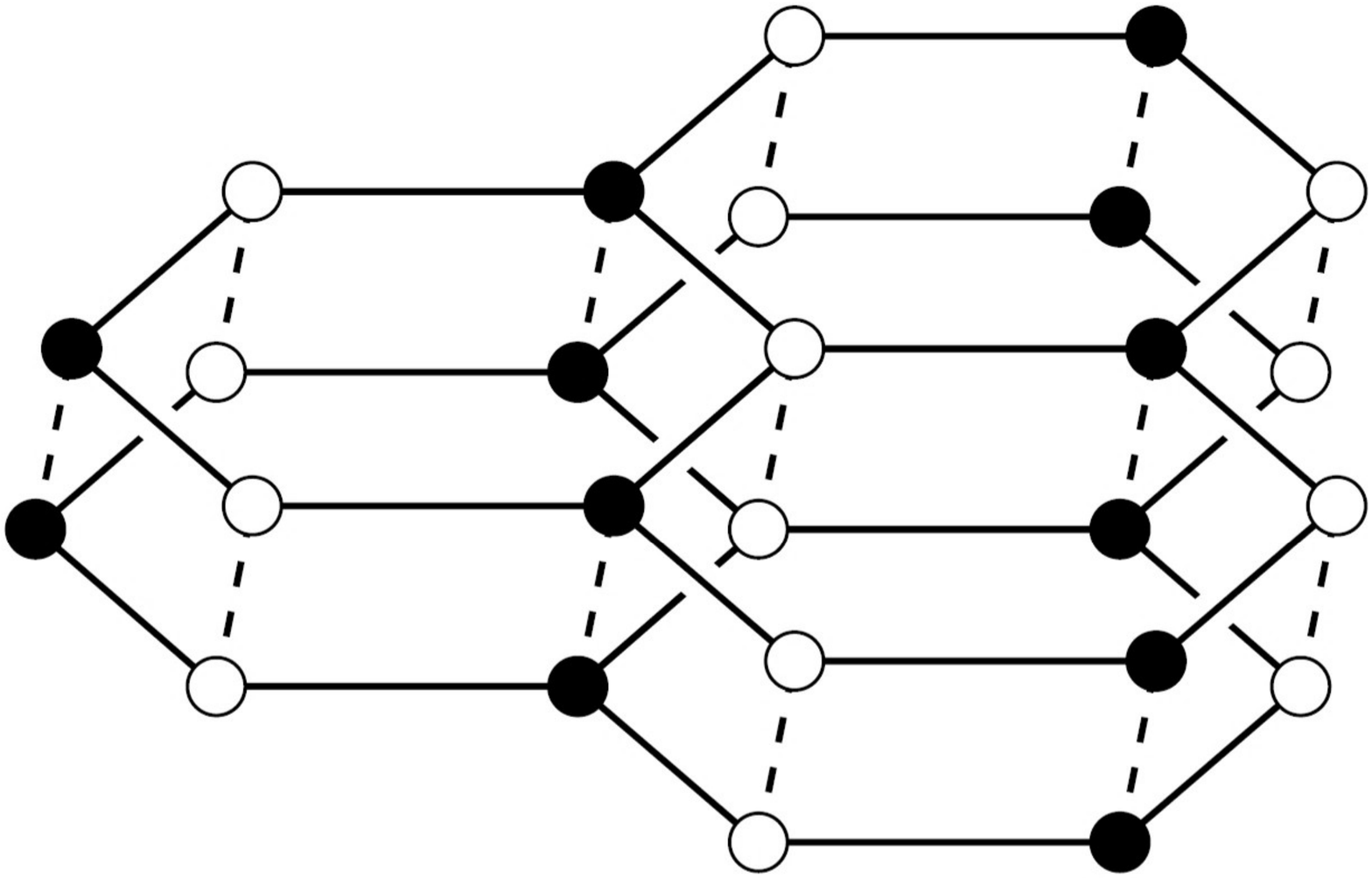}
\vspace{-0.5cm}
\caption{Ice with $1h$ structure.}
\label{FigAAAA}
\end{figure}

\noindent
In all the sequel, we shall focus our attention on two different type of structures. The first one corresponds to the $1h$ structure of the ice.
Sheets are stacked in such a way that the moving particle can always jump from one sheet to another one with small probability, as shown in Figure \ref{FigAAAA}.

One can observe that a particle located at a white vertex (resp. black vertex) is only allowed to jump to a white vertex (resp. black vertex).
The set of vertices are denoted once again by $\cV_0$ and $\cV_1$ where for $i=0,1$,
$$ \cV_i = \left\{ \Bigl(a\times\Bigl(i+ \frac{3}{2}k\Bigr), a\times\Bigl(\frac{\sqrt{3}}{2}k+\sqrt{3}\ell\Bigr), \ h\times n\Bigr) \ : \ k,\ell,n\in \dZ\right\} $$
where $a$ stands for the distance between adjacent vertices located in the same sheet
and $h$ stands for the distance between consecutive sheets of the ice.
The index $i=0$ if the vertex is white and the index $i=1$ if the vertex is black. 
\ \par
\begin{figure}[htbp] 
    \centering
    \vspace{-1.2cm}
\includegraphics[scale=0.32,angle=-90,origin=c]{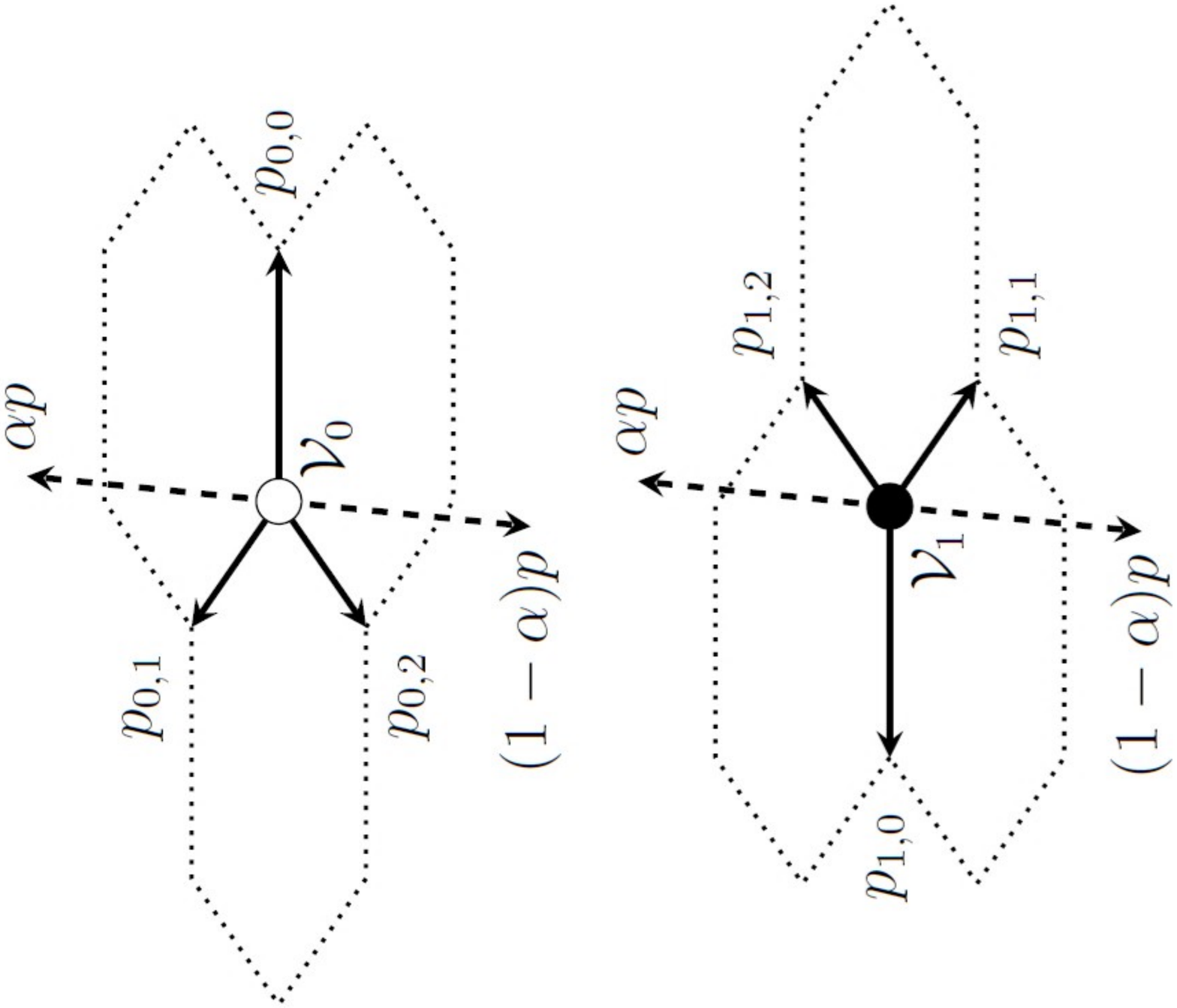}
\vspace{-1cm}
    \caption{Transition probabilities for the $1h$ structure of the ice.}
    \label{FigTransAAAA}
\end{figure}

The random walk on the ice with $1h$ structure is as follows.
At time zero, the particle starts at the origin $S_0=(0,0,0)$. Afterwards, at time $n \geq  0$, assume that the position of the
particle is given by $S_n=(X_n,Y_n,Z_n)$. Then, the particle can jump to an adjacent sheet with small probabilities, 
that is for $i=0,1$ and for all $(x,y,z) \in \cV_i$,
\begin{equation}
\label{TransAAV1}
    \dP \left[ S_{n+1} = \left(\begin{array}{c} x \\ y \\ z+h  \end{array}\right)
    \left\vert  \ S_n = \left(\begin{array}{c}x \\ y \\z \end{array}\right)\right.\right]=\alpha p
\end{equation}
while
\begin{equation}
\label{TransAAV2}
    \dP \left[ S_{n+1} = \left(\begin{array}{c} x \\ y \\ z-h  \end{array}\right)
    \left\vert  \ S_n = \left(\begin{array}{c}x \\ y \\z \end{array}\right)\right.\right]=(1-\alpha)p
\end{equation}
where $0\leqslant p\leqslant 1$ and $0<\alpha<1$, the symmetrical case corresponding to $\alpha=1/2$. 
Otherwise, if the particle remains on the same sheet, the transition probabilities are the same as those in (\onlinecite{Crescenzo19}), that is 
for $i=0,1$,  for all $(x,y,z) \in \cV_i$ and for $j=0,1,2$,
\begin{equation}
\label{TransAAH}
    \dP \left[ S_{n+1} = 
    \left(\begin{array}{c} x+a\cos \bigl(\frac{2}{3}j\pi + i\pi  \bigr) \\ y+a\sin \bigl(\frac{2}{3}j\pi + i\pi \bigr)\\z \end{array}\right)
    \left\vert \  S_n = \left(\begin{array}{c}x \\ y \\z \end{array}\right)\right.\right]=p_{i,j}
\end{equation}
where for $i=0,1$,
$$
\sum_{j=0}^{2} p_{i,j}=1-p.
$$
The transition probabilities are represented in Figure~\ref{FigTransAAAA}. More precisely, if the particle is located in a vertex of $\cV_0$,
it can jump to the sheets above or below in a vertex of $\cV_0$ with small probabilities $\alpha p$ and $(1-\alpha) p$ respectively, or
it can reach the three adjacent vertices of $\cV_1$ with probabilities $p_{0,0}$, $p_{0,1}$ and $p_{0,2}$.
By the same token, if the particle is located in a vertex of $\cV_1$,
it can jump to the sheets above or below in a vertex of $\cV_1$ with small probabilities $\alpha p$ and $(1-\alpha) p$ respectively, or
it can reach the three adjacent vertices of $\cV_0$ with probabilities $p_{1,0}$, $p_{1,1}$ and $p_{1,2}$.

\begin{figure}[htbp] 
    \centering
\includegraphics[scale=0.35,angle=-90,origin=c]{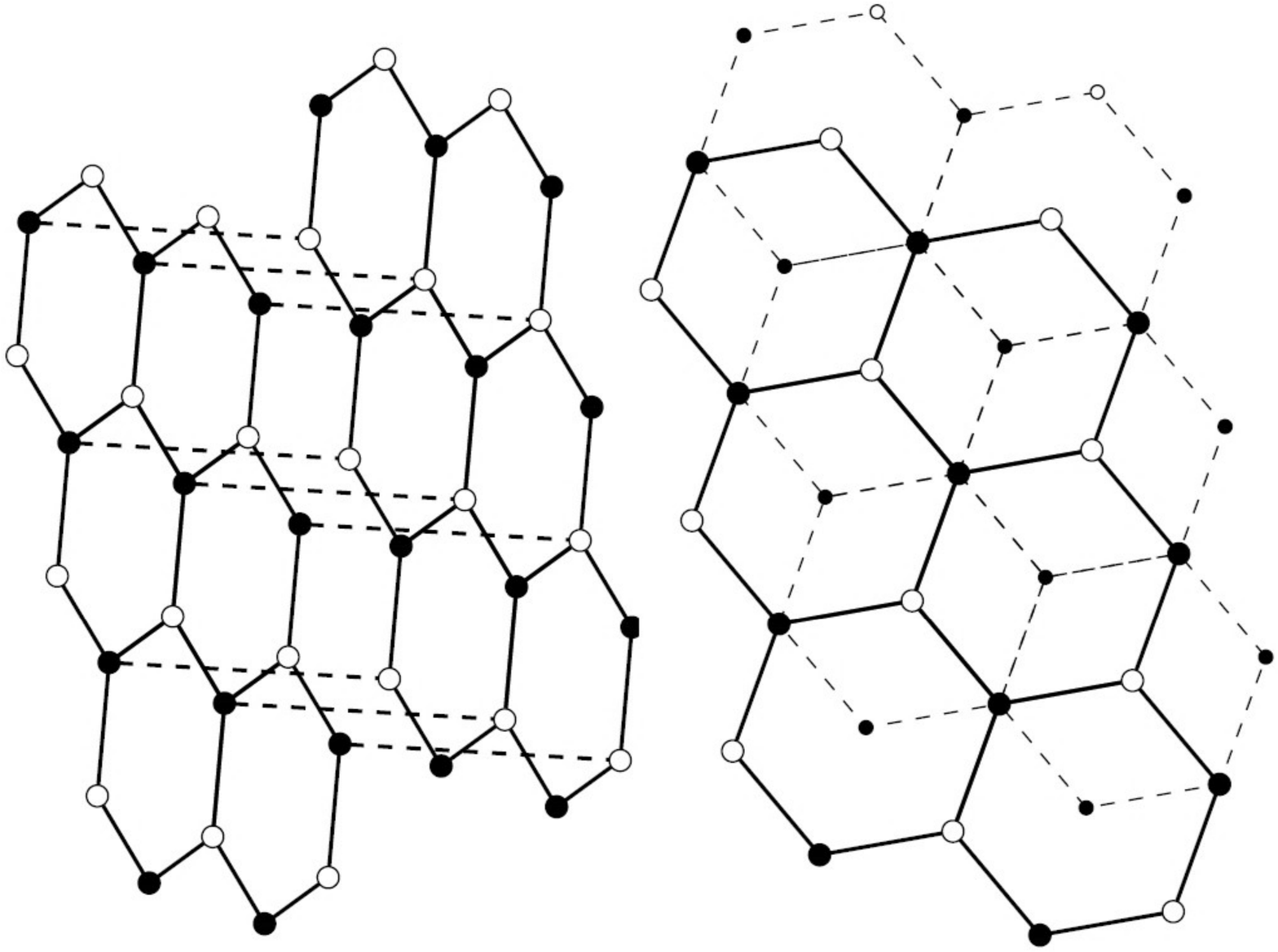}
\caption{Graphite with $2h$ structure.}
\label{FigABAB}
\end{figure}

%%%%%%%%%%%%%%%%%%%%%%%%%%%%%%%%%%%%%%%%%%%%%%%%%%%%%%%%%%%%%%%%%%%%%%%%%%%%%%%%%%%%%%%%%%%%%%%%%%%%%%%%%%%%%%%%%%%%%%%%
A second type of structure we are interested in, is the $2h$ structure of the graphite represented in Figure \ref{FigABAB} where a particle 
located at a white vertex (resp. black vertex) can only jump to a black vertex (resp. white vertex). In other words, white vertices 
(resp. black vertices) of a given sheet are only connected to black vertices
(resp. white vertices) of the graphite sheets just above or below.

\noindent
The set of vertices are now denoted by $\cV_{0,0}$, $\cV_{1,0}$ and $\cV_{0,1}$, $\cV_{1,1}$ where for $i=0,1$ and $j=0,1$,
\begin{align*}
\cV_{i,j} \!=\! \Bigg\{ \Biggl(a\times\Bigl((-1)^{i+1}& \rI_{j=1} +\frac{3}{2}k \Bigr), a\times\Bigl(\frac{\sqrt{3}}{2}k+\sqrt{3}\ell\Bigr), \\
&h\times (2n+ \rI_{i\neq j})\Biggr) 
\ \text{with} \ k,\ell,n\in \dZ\Bigg\} 
\end{align*} 
where as before $a$ is the distance between adjacent vertices located in the same sheet and $h$ is the distance between consecutive
sheets of graphene. The index $i=0$ if the vertex is white and $i=1$ if the vertex is black (which refers to the horizontal local neighborhood), while the index $j=0$ if the particle can move 
to an adjacent sheet from this vertex and $j=1$ otherwise. The main difference with the $1h$ structure of the ice is that here
the particle does not always have the possibility to jump to an adjacent sheet.
\ \par
\begin{figure*}[tbp] 
    \centering
    \vspace{-5cm}
\includegraphics[scale=0.6]{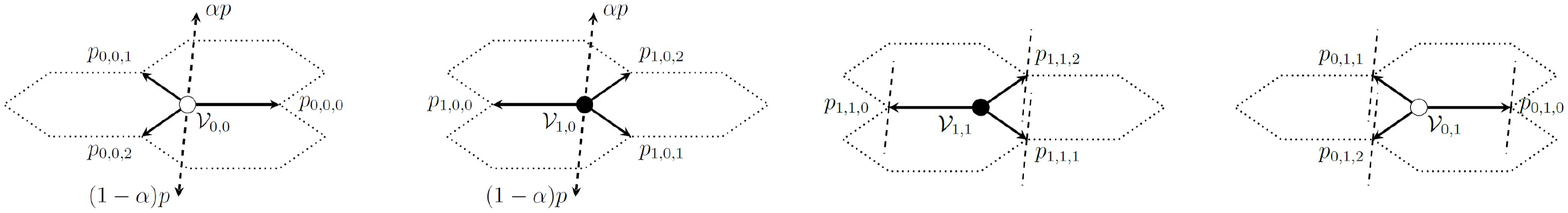}
 \vspace{-5cm}
    \caption{Transition probabilities for the $2h$ structure of the graphite.}
    \label{FigTransABAB}
\end{figure*}

The random walk on the graphite with $2h$ structure is as follows.
At time zero, the particle starts at the origin $S_0=(0,0,0)$. Afterwards, at time $n \geq  0$, assume that the position of the
particle is given by $S_n=(X_n,Y_n,Z_n)$. Then, for $i=0,1$, if the particle is located in $\cV_{i,0}$, it has the possibility to
jump to an adjacent sheet with small probabilities, that is for $i=0,1$ and for all $(x,y,z) \in \cV_{i,0}$,
\begin{equation}
\label{TransABV1}
    \dP \left[ S_{n+1} = \left(\begin{array}{c} x \\ y \\ z+h  \end{array}\right)
    \left\vert  \ S_n = \left(\begin{array}{c}x \\ y \\z \end{array}\right)\right.\right]=\alpha p
\end{equation}
while
\begin{equation}
\label{TransABV2}
    \dP \left[ S_{n+1} = \left(\begin{array}{c} x \\ y \\ z-h  \end{array}\right)
    \left\vert \ S_n= \left(\begin{array}{c}x \\ y \\z \end{array}\right)\right.\right]=(1-\alpha)p
\end{equation}
where $0\leqslant p\leqslant 1$ and $0<\alpha<1$, the symmetrical case corresponding to $\alpha=1/2$. 
Otherwise, for $i=0,1$, if the particle is located in $\cV_{i,0}$ and it remains on the same sheet, the transition probabilities 
are given for all $(x,y,z) \in \cV_{i,0}$ and for $k=0,1,2$, by
\begin{equation}
\label{TransABH}
    \dP \left[ S_{n+1} = 
    \left(\begin{array}{c} x+a\cos \bigl(\frac{2}{3}k\pi + i\pi  \bigr) \\ y+a\sin \bigl(\frac{2}{3}k\pi + i\pi \bigr)\\z \end{array}\right)
    \left\vert \ S_n = \left(\begin{array}{c}x \\ y \\z \end{array}\right)\right.\right]=p_{i,0,k}
\end{equation}
where for $i=0,1$,
$$
\sum_{k=0}^{2} p_{i,0,k}=1-p.
$$
Finally, for $i=0,1$, if the particle is located in $\cV_{i,1}$, the transition probabilities are the same as those in (\onlinecite{Crescenzo19}), that is 
for $i=0,1$,  for all $(x,y,z) \in \cV_{i,1}$ and for $k=0,1,2$,
\begin{equation*}
    \dP \left[ S_{n+1} = 
    \left(\begin{array}{c} x+a\cos \bigl(\frac{2}{3}k\pi + i\pi  \bigr) \\ y+a\sin \bigl(\frac{2}{3}k\pi + i\pi \bigr)\\z \end{array}\right)
    \left\vert \ S_n = \left(\begin{array}{c}x \\ y \\z \end{array}\right)\right.\right]=p_{i,1,k}
\end{equation*} 
where for $i=0,1$,
$$
\sum_{k=0}^{2} p_{i,1,k}=1.
$$
As it was previously done for the $1h$ structure of the ice, the transition probabilities for the $2h$ structure
of the graphite are given in Figure~\ref{FigTransABAB}.

\begin{figure}[htbp] 
\centering
\includegraphics[scale=0.28]{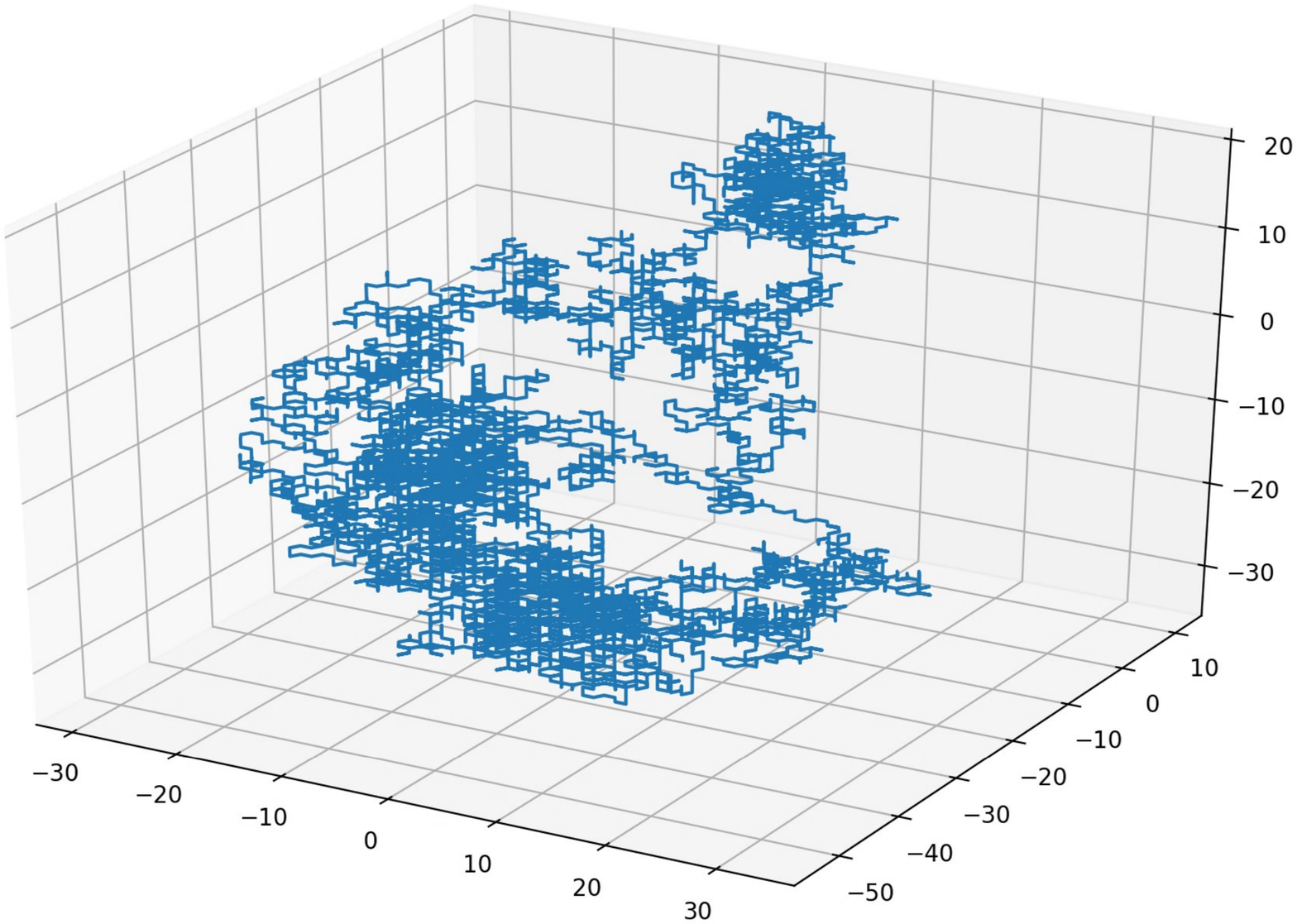}
\caption{Trajectory of the three-dimensional RWI.}
\label{FigTrajectoryAAAA}
\end{figure}

\begin{figure}[htbp] 
\centering
\includegraphics[scale=0.28]{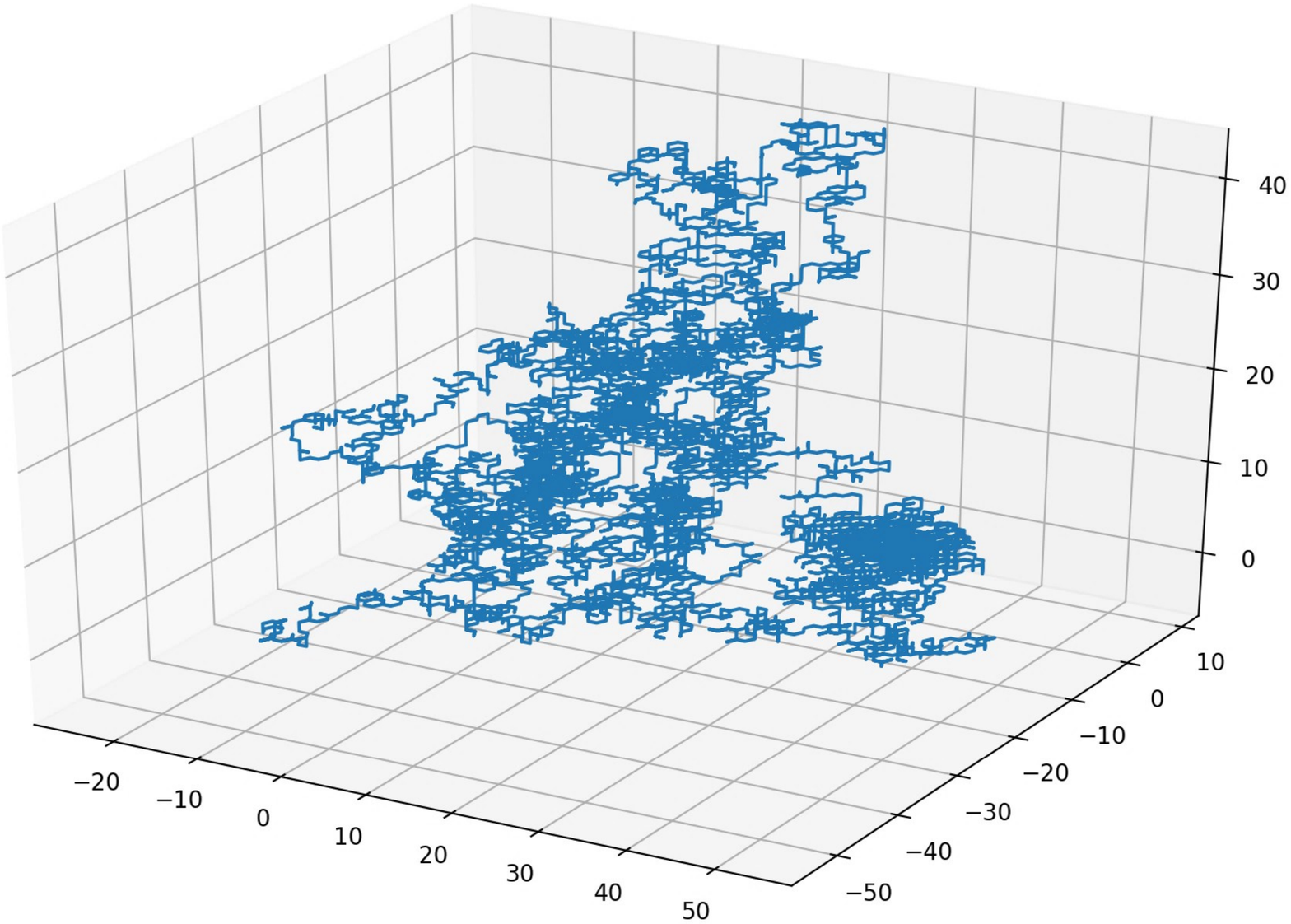}
\caption{Trajectory of the three-dimensional RWG.}
\label{FigTrajectoryABAB}
\end{figure}

The goal of this paper is to investigate the asymptotic behavior of three-dimensional RWI and RWG
with this two different type of structures. 
Figures ~\ref{FigTrajectoryAAAA} and ~\ref{FigTrajectoryABAB} shows two trajectories of length $n=10\,000$ of the RWI and RWG, respectively. 
The distance $a$ between adjacent vertices and the distance $h$ between consecutive sheets are given by $a=1$ and $h=1$, 
while the probability to jump to an adjacent sheet $p=1/5$, $\alpha=1/2$ and the transitions probabilities 
are given, for $i=0,1$ and $j=0,1,2$, by

$$
p_{i,j}=\frac{1}{3}(1-p).
$$


%%%%%%%%%%%%%%%%%%%%%%%%%%%%%%%%%%%%%%%%%%%%%%%%%%%%%%%%%%%%%%%%%%%%%%%%%%%%%%%%%%%%%%%%%%%%%%%%%%%%%%%%%%%%%%%%%%%%%%%%

\section{Main results}
\label{S-MR}
%%%%%%%%%%%%%%%%%%%%%%%%%%%%%%%%%%%%%%%%%%%%%%%%%%%%%%%%%%%%%%%%%%%%%%%%%%%%%%%%%%%%%%%%%%%%%%%%%%%%%%%%%%%%%%%%%%%%%%%%

Our first result concerns the strong law of large numbers for the random walk on ice structure. Let $\mu$ be the mean vector defined by
\begin{equation}
\label{DEFMUA}
\mu = \begin{pmatrix}
\mu_1 \\
\mu_2 \\
\mu_3
\end{pmatrix}
\end{equation}
with
\begin{eqnarray*}
\mu_1 &=&\frac{3a}{4}\Bigl(-u_0 +u_1\Bigr), \\
\mu_2 &=&\frac{a\sqrt{3}}{4}\Bigl(v_0-v_1\Bigr), \\
\mu_3 &=& \Big.hp(2\alpha -1),
\end{eqnarray*}
where for $i=0,1$, $u_i= p_{i,1} + p_{i,2}$ and $v_i=p_{i,1} - p_{i,2}$.

%%%%%%%%%%%%%%%%%%%%%%%%%%%%%%%%%%%%%%%%%%%%%%%%%%%%%%%%%%%%%%%%%%%%%%%%%%%%%%%%%%%%%%%%%%%%%%%%%%

\begin{thm}
\label{T-ASCVGRWGA}
For the RWI, we have the almost sure convergence
\begin{equation}
\label{ASCVGRWGA}
 \lim_{n \rightarrow \infty} \frac{1}{n}S_n=\mu \hspace{1cm} \text{a.s.}
\end{equation}
More precisely, 
\begin{equation}
\label{RATECVGRWGA}
\Bigl\| \frac{1}{n}S_n - \mu \Bigr\|^2=O\Bigl( \frac{\log n}{n} \Bigr)\hspace{1cm}\text{a.s.}
\end{equation}
\end{thm}

\noindent
Our second result is devoted to the asymptotic normality for the random walk on ice structure. For this purpose, 
denote
\begin{equation}
\label{DEFSIGMA2A}
\sigma^2 = \begin{pmatrix}
\sigma^2_1 & \sigma_{1,2} & 0\\
\sigma_{1,2} & \sigma^2_2 & 0 \\
0 & 0 & \sigma^2_3
\end{pmatrix}
\end{equation}
where 
\begin{align*}
\sigma^2_1 &=a^2 \Bigl(p(1-p)+\frac{3}{8}(3-4p)\bigl(u_0+u_1\bigr)
-\frac{9}{8}\bigl(u_0^2+u_1^2\bigr)\Bigr), \\
\sigma^2_2 &=\frac{3a^2}{8}\Bigl(\bigl(u_0+u_1\bigr)-\bigl(v_0^2+v_1^2\bigr)
\Bigr), \\
\sigma^2_3&= h^2p\Bigl(1-p\bigl(2\alpha-1\bigr)^2\Bigr), \\
\sigma_{1,2} &=\frac{a^2\sqrt{3}}{8}\Bigl(\bigl(-3+2p\bigr)\bigl(v_0+v_1\bigr) +3\bigl(u_0v_0+u_1v_1\bigr)\Bigr). 
\end{align*}
In addition, let
\begin{equation}
\label{DEFTHETAA}
\theta = \begin{pmatrix}
\theta_1 \\
\theta_2 \\
0
\end{pmatrix}
\end{equation}
with
\begin{eqnarray*}
\theta_1 &=&a\Bigl((1-p) -\frac{3}{4}(u_0+u_1)\Bigr), \\
\theta_2 &=&\frac{\sqrt{3}a}{4}\bigl(v_0+v_1\bigr). \\
\end{eqnarray*}

\begin{thm}
\label{T-ANRWGA}
For the RWI, we have the asymptotic normality
\begin{equation}
\label{ANRWGA}
\frac{1}{\sqrt{n}}\bigl( S_n -n \mu \bigr) \liml \cN \left(0, \Gamma \right)
\end{equation}
where the covariance matrix $\Gamma$ is given by $\Gamma=\sigma^2$ if $p=1$, whereas if
$0\leq p <1$,
\begin{equation}
\label{DEFGAMMAA}
\Gamma= \sigma^2 - \Bigl( \frac{p}{1-p}\Bigr)\theta \theta^T.
\end{equation}
\end{thm}

\noindent
Our third result deals with the strong law of large numbers for the random walk on graphite structure.
Denote by $\mu$ and $m$ the mean vectors 
\begin{equation}
\label{DEFMUTHETAB}
\mu = \begin{pmatrix}
\mu_1 \\
\mu_2 \\
\mu_3
\end{pmatrix}
\hspace{1cm}\text{and}\hspace{1cm}
m = \begin{pmatrix}
m_1 \\
m_2 \\
m_3
\end{pmatrix}
\end{equation}
with
\begin{eqnarray*}
\mu_1 &=&\frac{3a}{8}\Bigl(-(u_{0,0}+u_{0,1})+(u_{1,0}+u_{1,1})\Bigr), \\
\mu_2 &=&\frac{a\sqrt{3}}{8}\Bigl((v_{0,0}+v_{0,1})-(v_{1,0}+v_{1,1})\Bigr), \\
\mu_3 &=& \frac{hp}{2}(2\alpha -1)
\end{eqnarray*}
and
\begin{eqnarray*}
m_1 &=&\frac{3a}{8}\Bigl(-(u_{0,0}-u_{0,1})+(u_{1,0}-u_{1,1})\Bigr), \\
m_2 &=&\frac{a\sqrt{3}}{8}\Bigl((v_{0,0}-v_{0,1})-(v_{1,0}-v_{1,1})\Bigr), \\
m_3 &=& \frac{hp}{2}(2\alpha -1)
\end{eqnarray*}
where for $i,j=0,1$, $u_{i,j}=p_{i,j,1}+p_{i,j,2}$ and $v_{i,j}=p_{i,j,1}-p_{i,j,2}$.
For the sake of clarity, we have chosen to keep the same notation for the mean vector $\mu$  
in both hexagonal structures. We shall also make use of the vectors 
$\theta$ and $\rho$ defined by
\begin{equation}
\label{DEFRHOTHETAB}
\theta = \begin{pmatrix}
\theta_1 \\
\theta_2 \\
0
\end{pmatrix}
\hspace{1cm}\text{and}\hspace{1cm}
\rho = \begin{pmatrix}
\rho_1 \\
\rho_2 \\
0
\end{pmatrix}
\end{equation}
with
\begin{eqnarray*}
\theta_1 &=& a\Bigl( \Bigl( 1-\frac{p}{2}\Bigr) - \frac{3}{8}\Bigl( (u_{0,0}+u_{0,1})+(u_{1,0} +u_{1,1})\Bigr) \Bigr), \\
\theta_2 &=&\frac{a\sqrt{3}}{8}\Bigl((v_{0,0}+v_{0,1})+(v_{1,0}+v_{1,1})\Bigr), \\
\rho_1 &=& a\Bigl( -\frac{p}{2} - \frac{3}{8}\Bigl( (u_{0,0}-u_{0,1})+(u_{1,0} -u_{1,1})\Bigr) \Bigr), \\
\rho_2 &=&\frac{a\sqrt{3}}{8}\Bigl((v_{0,0}-v_{0,1})+(v_{1,0}-v_{1,1})\Bigr).
\end{eqnarray*}

\begin{thm}
\label{T-ASCVGRWGB}
For the RWG with $p>0$, we have the almost sure convergence
\begin{equation}
\label{ASCVGRWGB}
 \lim_{n \rightarrow \infty} \frac{1}{n}S_n=\mu + \Bigl(\frac{p}{2-p}\Bigr)m\hspace{1cm} \text{a.s.}
\end{equation}
More precisely, 
\begin{equation}
\label{RATECVGRWGB}
\Bigl\| \frac{1}{n}S_n - \mu - \Bigl( \frac{p}{2-p}\Bigr) m \Bigr\|^2=O\Bigl( \frac{\log n}{n} \Bigr)\hspace{1cm}\text{a.s.}
\end{equation}
\end{thm}

\begin{rem}
In the special case where $p=0$, the limiting value in \eqref{ASCVGRWGB} and \eqref{RATECVGRWGB} 
changes to $\mu + \rho$,
\begin{equation}
\label{ASCVGRWGBP0}
\lim_{n \rightarrow \infty} \frac{1}{n}S_n= \mu + \rho \hspace{1cm}\text{a.s.}
\end{equation}
and
\begin{equation}
\label{RATECVGRWGBP0}
\Bigl\| \frac{1}{n}S_n - \mu - \rho \Bigr\|^2=O\Bigl( \frac{\log n}{n} \Bigr)\hspace{1cm}\text{a.s.}
\end{equation}
If we denote $u_0=u_{0,0}$, $u_1=u_{1,1}$ and $v_0=v_{0,0}$, $v_1=v_{1,1}$, one can immediately see that
the almost sure convergences \eqref{ASCVGRWGA} and \eqref{ASCVGRWGBP0} are of course the same in the case $p=0$.
\end{rem}

\noindent
Our fourth result is dedicated to the asymptotic normality for the random walk on the graphene. To this end, let
\begin{equation}
\label{DEFZETA}
\zeta=\begin{pmatrix}
0\\
0 \\
hp(2\alpha-1)
\end{pmatrix}.
\end{equation}
Moreover, denote
\begin{equation}
\label{DEFALPHAGAMMA}
\sigma^2 = \begin{pmatrix}
\sigma^2_1 & \sigma_{1,2} & \sigma_{1,3}\\
\sigma_{1,2} & \sigma^2_2 & \sigma_{2,3} \\
\sigma_{1,3} & \sigma_{2,3} & \sigma^2_3
\end{pmatrix}
\hspace{1cm}\text{and}\hspace{1cm}
\gamma = \begin{pmatrix}
\gamma_1 & \gamma_4 & \gamma_6\\
\gamma_4 & \gamma_2 & \gamma_5 \\
\gamma_6 & \gamma_5 & \gamma_3
\end{pmatrix}
\end{equation}
where 
\begin{align*}
\sigma^2_1 &= \frac{a^2}{4}\Bigl(2p(1-p)+\frac{9}{4}\bigl(s_{0,0}+s_{1,0}+s_{0,1}+s_{1,1}\bigr)-
3p\bigl(u_{0,0}+u_{1,0}\bigr)\Bigr), \\
\sigma^2_2 &= \frac{3a^2}{16}\Bigl( \bigl(u_{0,0}+u_{1,0}+u_{0,1}+u_{1,1}\bigr) - \bigl(v_{0,0}^2+v_{1,0}^2+v_{0,1}^2+v_{1,1}^2\bigr)\Bigr), \\
\sigma^2_3 &= \frac{h^2 p}{2}\Bigl(1-p(2\alpha-1)^2\Bigr), \\
\sigma_{1,2} &= \frac{a^2\sqrt{3}}{16}\Bigl(3\bigl(t_{0,0}+t_{1,0}+t_{0,1}+t_{1,1}\bigr) +2p\bigl(v_{0,0}+v_{1,0}\bigr)\Bigr), \\
\sigma_{2,3} &= \frac{-ah \sqrt{3}}{8}\Bigl(p(2\alpha-1)\bigl(v_{0,0}+v_{1,0}\bigr)\Bigr),\\
\sigma_{1,3} &= \frac{ah}{8}\Bigl(-4p(1-p)(2\alpha-1)+3p(2\alpha-1)\bigl(u_{0,0}+u_{1,0}\bigr)\Bigr), 
\end{align*}
and
\begin{align*}
\gamma_1 &= \frac{a^2}{4}\Bigl(2p(1-p)+\frac{9}{4}\bigl(s_{0,0}+s_{1,0}-s_{0,1}-s_{1,1}\bigr)-
3p\bigl(u_{0,0}+u_{1,0}\bigr)\Bigr), \\
\gamma_2 &= \frac{3a^2}{16}\Bigl( \bigl(u_{0,0}+u_{1,0}-u_{0,1}-u_{1,1}\bigr) - \bigl(v_{0,0}^2+v_{1,0}^2-v_{0,1}^2-v_{1,1}^2\bigr)\Bigr), \\
\gamma_3 &= \frac{h^2 p}{2}\Bigl(1-p(2\alpha-1)^2\Bigr), \\
\gamma_4 &= \frac{a^2 \sqrt{3}}{16}\Bigl(3\bigl(t_{0,0}+t_{1,0}-t_{0,1}-t_{1,1}\bigr) +2p\bigl(v_{0,0}+v_{1,0}\bigr)\Bigr), \\
\gamma_5 &= \frac{-ah \sqrt{3}}{8}\Bigl(p(2\alpha-1)\bigl(v_{0,0}+v_{1,0}\bigr)\Bigr),\\
\gamma_6 &= \frac{ah}{8}\Bigl(-4p(1-p)(2\alpha-1)+3p(2\alpha-1)\bigl(u_{0,0}+u_{1,0}\bigr)\Bigr), 
\end{align*}
with for all $i=0,1$ and $j=0,1$, $s_{i,j}=u_{i,j}(1-u_{i,j})$ and ${t_{i,j}=v_{i,j}(u_{i,j}-1)}$.

\begin{thm}
\label{T-ANRWGB}
For the RWG with $p>0$, we have the asymptotic normality
\begin{equation}
\label{ANRWGB}
\frac{1}{\sqrt{n}}\bigl( S_n -n \mu - n\Bigl( \frac{p}{2-p}\Bigr) m \bigr) \liml \cN \left(0, \Gamma \right)
\end{equation}
where the covariance matrix $\Gamma$ is given by 
\begin{align}
\Gamma &=\sigma^2+ \frac{p}{2-p}\gamma +
   \frac{2}{(2-p)^2}\Bigl(\bigl(\zeta-p\mu\bigr)m^T + m\bigl(\zeta-p\mu\bigr)^T\Bigr) \nonumber\\
   & \hspace{0.5cm} -\frac{4p}{(2-p)^3}m m^T  + \frac{2}{2-p}\bigl(\theta \rho^T + \rho \theta^T\bigr) + \frac{4}{p(2-p)}\rho\rho^T.
\label{DEFGAMMAB}
\end{align}
\end{thm}

\begin{rem}
In the special case where $p=0$, we find that
\begin{equation}
\label{ANRWGBP0}
\frac{1}{\sqrt{n}}\bigl( S_n - n (\mu + \rho) \bigr)  \liml \cN \left(0, \sigma^2+\delta \right)
\end{equation}
where the matrix $\delta$ is defined in \eqref{DEFNUDELTA}. One can also observe that
the asymptotic variances in \eqref{ANRWGA} and \eqref{ANRWGBP0} coincide in the case $p=0$.
\end{rem}

%%%%%%%%%%%%%%%%%%%%%%%%%%%%%%%%%%%%%%%%%%%%%%%%%%%%%%%%%%%%%%%%%%%%%%%%%%%%%%%%%%%%%%%%%%%%%%%%%%%%%%%%%%%%%%%%%%%%%%%

\section{Proofs for the RWI}
\label{S-PRAA}
%%%%%%%%%%%%%%%%%%%%%%%%%%%%%%%%%%%%%%%%%%%%%%%%%%%%%%%%%%%%%%%%%%%%%%%%%%%%%%%%%%%%%%%%%%%%%%%%%%%%%%%%%%%%%%%%%%%%%%%%

%%%%%%%%%%%%%%%%%%%%%%%%%%%%%%%%%%%%%%%%%%%%%%%%%%%%%%%%%%%%%%%%%%%%%%%%%%%%%%%%%%%%%%%%%%%%%%%%%%%%%%%%%%%%%%%%%%%%%%%%

\noindent{\bf Proof of Theorem \ref{T-ASCVGRWGA}.}
%%%%%%%%%%%%%%%%%%%%%%%%%%%%%%%%%%%%%%%%%%%%%%%%%%%%%%%%%%%%%%%%%%%%%%%%%%%%%%%%%%%%%%%%%%%%%%%%%%%%%%%%%%%%%%%%%%%%%%%%
In order to prove the almost sure convergence \eqref{ASCVGRWGA}, denote by $(\xi_n)$ the increments of the RWI.
Then, the position of the RWI is given, for all $n \geq 0$, by
\begin{equation}
\label{POSRWI}
S_{n+1}=S_n+\xi_{n+1}= \sum_{k=1}^{n+1} \xi_k
\end{equation}
where
$$
\xi_{n+1}=\begin{pmatrix}
X_{n+1}-X_n \\
Y_{n+1}-Y_n\\
Z_{n+1}-Z_n
\end{pmatrix}.
$$
Let $(\cF_n)$ be the natural filtration associated with the RWI, that is $\cF_n=\sigma\big(\xi_1,\dots,\xi_n\big)$.
We have for all $n\geq 0$,
$\dE[ \xi_{n+1} | \cF_n] = \dE[ \xi_{n+1} \vert S_n\in\cV_0]\rI_{S_n\in\cV_0} + 
\dE[ \xi_{n+1} | S_n\in\cV_1]\rI_{S_n\in\cV_1}$.
Hence, it follows from \eqref{TransAAV1}, \eqref{TransAAV2} and \eqref{TransAAH} that
\begin{align*}
    \dE [ \xi_{n+1} | \cF_n] \!=\! 
    \begin{pmatrix}
    a\bigl( p_{0,0} -\frac{1}{2}( p_{0,1} + p_{0,2}) \bigr)\\
    \frac{a \sqrt{3}}{2}\bigl( p_{0,1} - p_{0,2} \bigr)\\
    hp(2\alpha-1)
    \end{pmatrix}\!\rI_{S_n\in\cV_0} &\\
    \!+\!
    \begin{pmatrix}
    a\bigl( -p_{1,0} +\frac{1}{2}( p_{1,1} + p_{1,2} )\bigr)\\
    \frac{a \sqrt{3}}{2}\bigl( p_{1,2} - p_{1,1} \bigr)\\
    hp(2\alpha-1)
    \end{pmatrix}\!\rI_{S_n\in\cV_1}&.
\end{align*}
For $i=0,1$, denote $u_i= p_{i,1} + p_{i,2}$ and $v_i= p_{i,1} - p_{i,2}$. We clearly have for $i=0,1$, $p_{i,0}=1-p-u_{i}$. Consequently,
$\dE [ \xi_{n+1} | \cF_n]$ reduces to
\begin{equation}
\label{CEXIAA}
 \dE [ \xi_{n+1} | \cF_n]= (\mu + \theta)\rI_{S_n\in\cV_0}+(\mu - \theta)\rI_{S_n\in\cV_1}= \mu + \theta \varepsilon_n
\end{equation}
where $\varepsilon_n$ stands for the random variable 
\begin{equation}
\label{DEFVAREPSILON}
\varepsilon_n=\rI_{S_n\in\cV_0}-\rI_{S_n\in\cV_1}
\end{equation}
and the vectors $\mu$ and $\theta$ are given by \eqref{DEFMUA} and \eqref{DEFTHETAA}.
Moreover, we have for all $n\geq 0$,
$\dE[ \xi_{n+1}\xi_{n+1}^T | \cF_n] = \dE[ \xi_{n+1}\xi_{n+1}^T \vert S_n\in\cV_0]\rI_{S_n\in\cV_0} + 
\dE[ \xi_{n+1}\xi_{n+1}^T | S_n\in\cV_1]\rI_{S_n\in\cV_1}$.
We obtain once again from \eqref{TransAAV1}, \eqref{TransAAV2} and \eqref{TransAAH} that
\begin{align*}
    &\dE [ \xi_{n+1}\xi_{n+1}^T | \cF_n] = \\
    &\begin{pmatrix}
    a^2\bigl( p_{0,0} +\frac{1}{4}( p_{0,1} + p_{0,2}) \bigr) &  -\frac{a^2 \sqrt{3}}{4}\bigl( p_{0,1} - p_{0,2} \bigr) & 0\\
    -\frac{a^2 \sqrt{3}}{4}\bigl( p_{0,1} - p_{0,2} \bigr) & \frac{3 a^2}{4}\bigl( p_{0,1} + p_{0,2} \bigr) & 0\\
    0 & 0 & h^2 p
    \end{pmatrix}\rI_{S_n\in\cV_0} 
    \\
    & +
    \begin{pmatrix}
    a^2\bigl( p_{1,0} +\frac{1}{4}( p_{1,1} + p_{1,2}) \bigr) &  -\frac{a^2 \sqrt{3}}{4}\bigl( p_{1,1} - p_{1,2} \bigr) & 0\\
    -\frac{a^2 \sqrt{3}}{4}\bigl( p_{1,1} - p_{1,2} \bigr) & \frac{3 a^2}{4}\bigl( p_{1,1} + p_{1,2} \bigr) & 0\\
    0 & 0 & h^2 p
    \end{pmatrix}\rI_{S_n\in\cV_1}.
\end{align*}
It implies that
\begin{align}
   & \dE[\xi_{n+1}\xi_{n+1}^T | \cF_n] = 
    \Bigl(\sigma^2+\nu+(\mu+\theta)(\mu+\theta)^T\Bigr)\rI_{S_n\in\cV_0} \nonumber \\
    &\hspace{2.3cm}+ \Bigl(\sigma^2-\nu+(\mu-\theta)(\mu-\theta)^T\Bigr)\rI_{S_n\in\cV_1}, \nonumber \\
    &= \Bigl(\sigma^2 + \mu\mu^T + \theta\theta^T\Bigr) +   \Bigl(\nu+\mu\theta^T + \theta\mu^T\Bigr)\varepsilon_n
\label{CEXI2AA}
\end{align}
where the covariance matrix $\sigma^2$ and the random variable $\varepsilon_n$  are respectively given by \eqref{DEFSIGMA2A} and \eqref{DEFVAREPSILON}, while the deterministic matrix $\nu$ is defined by
$$
\nu = \begin{pmatrix}
\nu_1 & \nu_3 & 0\\
\nu_3 & \nu_2 & 0 \\
0 & 0 & 0
\end{pmatrix}
$$
with
\begin{align*}
\nu_1 &=\frac{3a^2}{8} \bigl(u_0-u_1\bigr)
\Bigl(3-4p -3 \bigl(u_0+u_1\bigr)\Bigr), \\
\nu_2 &=\frac{3a^2}{8} \Bigl( u_0-u_1-v_0^2+v_1^2\Bigr), \\
\nu_3 &=\frac{a^2\sqrt{3}}{8}\Bigl(\bigl(-3+2p\bigr)\bigl(v_0-v_1\bigr) +3\bigl(u_0v_0-u_1v_1\bigr)\Bigr).
\end{align*}
Hereafter, we have the martingale decomposition
\begin{equation}
\label{DECMARTAA}
S_n = \sum_{k=1}^n \bigl(\xi_k-\dE[\xi_k | \cF_{k-1}] \bigr)+\sum_{k=1}^n\dE[\xi_k| \cF_{k-1}] = M_n + R_n 
\end{equation}
where $(M_n)$ is the locally square integrable martingale given by
\begin{equation}
\label{DEFMARTAA}
M_n= \sum_{k=1}^n \bigl(\xi_k-\dE[\xi_k | \cF_{k-1}]\bigr)
\end{equation}
and the centering term
\begin{equation}
\label{DEFRMAA}
R_n=\sum_{k=1}^n\dE[\xi_k | \cF_{k-1}].
\end{equation}
The predictable quadratic variation \cite{Duflo97} associated with
$(M_n)$ is the random matrix given, for all $n \geq 1$, by
$$
\langle M\rangle_n = \sum_{k=1}^n \bigl(\dE[\xi_k\xi_k^T | \cF_{k-1}]-\dE[\xi_k| \cF_{k-1}]\dE[\xi_k| \cF_{k-1}]^T\bigr).
$$
It follows from \eqref{CEXIAA} and \eqref{CEXI2AA} that
\begin{align*}
\langle M\rangle_n = \sum_{k=1}^n \bigl((\sigma^2 + \mu\mu^T + \theta\theta^T) +   (\nu+\mu\theta^T + \theta\mu^T)\varepsilon_{k-1} &\\
- ( \mu + \theta \varepsilon_{k-1}) ( \mu + \theta \varepsilon_{k-1})^T \bigr)&
\end{align*}
which reduces to
\begin{equation}
\langle M\rangle_n= n \sigma^2 +I_{n-1} \nu 
\label{IPMN}
\end{equation}
where
$$
I_n=\sum_{k=0}^n \varepsilon_{k}.
$$
Furthermore, we clearly have for all $n \geq 0$,
\begin{align} 
\label{CEVAREPSAA}
   \dE[\varepsilon _{n+1} | \cF_n] & = \dE[ \varepsilon_{n+1} | S_n\in\cV_0]\rI_{S_n\in\cV_0} + 
   \dE[ \varepsilon_{n+1} | S_n\in\cV_1]\rI_{S_n\in\cV_1}, \nonumber \\
    & = (2p-1) \rI_{S_n\in\cV_0} + (1-2p) \rI_{S_n\in\cV_1}, \nonumber \\
    & = (2p-1) \varepsilon_n.
\end{align}
Consequently, we obtain the second martingale decomposition
\begin{align}
I_n & = \sum_{k=0}^n \varepsilon_k= 1+ \sum_{k=1}^n \bigl(\varepsilon_k-\dE[\varepsilon_k | \cF_{k-1}] \bigr)+\sum_{k=1}^n\dE[\varepsilon_k| \cF_{k-1}],
\nonumber \\
 & = 1+N_n + \sum_{k=1}^n (2p-1) \varepsilon_{k-1}=1+N_n+(2p-1) I_{n-1}
\label{DECMARTNEWAA}
\end{align}
where $N_n$ is the locally square integrable martingale given by
\begin{equation}
\label{DEFMARTNEWAA}
N_n= \sum_{k=1}^n \bigl(\varepsilon_k-\dE[\varepsilon_k | \cF_{k-1}]\bigr).
\end{equation}
We deduce from \eqref{CEVAREPSAA} that the predictable quadratic variation associated with
$(N_n)$ is  given by
\begin{align}
\langle N\rangle_n &= \sum_{k=1}^n \bigl(\dE[\varepsilon_k^2| \cF_{k-1}]-\dE^2[\varepsilon_k| \cF_{k-1}]\bigr) \nonumber \\
&=n-\sum_{k=1}^n (2p-1)^2\varepsilon_{k-1}^2 =4p(1-p)n.
\label{IPNN}
\end{align}
Therefore, we immediately obtain that
\begin{equation}
\label{CVGIPNN}
\lim_{n \rightarrow \infty} \frac{1}{n}\langle N\rangle_n= 4p(1-p) \hspace{1cm}\text{a.s.}
\end{equation}
One can also observe that the increments of the martingale $(N_n)$ are bounded by 2.
Hence, by virtue of the strong law of large numbers for martingales 
\begin{equation}
\label{CVGMGNN}
\lim_{n \rightarrow \infty} \frac{1}{n}N_n= 0 \hspace{1cm}\text{a.s.}
\end{equation}
More precisely, it follows from the last part of Theorem 1.3.24 in (\onlinecite{Duflo97}) that
\begin{equation}
\label{RATECVGMGNN}
N_n^2=O(n \log n) \hspace{1cm}\text{a.s.}
\end{equation}
However, we infer from \eqref{DECMARTNEWAA} together with the definition of $I_n$ that
$$
2(1-p)I_n=1+N_n -(2p-1) \varepsilon_n.
$$
Consequently, if $p<1$, we obtain from \eqref{CVGMGNN} that
\begin{equation}
\label{CVGLN}
\lim_{n \rightarrow \infty} \frac{1}{n}I_n= 0 \hspace{1cm}\text{a.s.}
\end{equation}
In addition, \eqref{RATECVGMGNN} clearly leads to
\begin{equation}
\label{RATECVGLN}
I_n^2=O(n \log n) \hspace{1cm}\text{a.s.}
\end{equation}
Then, we find from \eqref{IPMN} together with \eqref{CVGLN}  that
\begin{equation}
\label{CVGIPMN}
\lim_{n \rightarrow \infty} \frac{1}{n}\langle M\rangle_n= \sigma^2\hspace{1cm}\text{a.s.}
\end{equation}
In the special case where $p=1$, we easily see that $\theta =0$ and $\nu=0$, thus \eqref{CVGIPMN} still holds.
Therefore, in both cases, we obtain from the strong law of large numbers for martingales that
\begin{equation}
\label{CVGMGMN}
\lim_{n \rightarrow \infty} \frac{1}{n}M_n= 0 \hspace{1cm}\text{a.s.}
\end{equation}
More precisely, one can observe that the increments of $(M_n)$ are almost surely bounded.
Hence, by examining each component of the martingale $(M_n)$, it follows once again from the last part of 
Theorem 1.3.24 in (\onlinecite{Duflo97}) that
\begin{equation}
\label{RATECVGMGNM}
||M_n||^2=O(n \log n) \hspace{1cm}\text{a.s.}
\end{equation}
The centering term $R_n$ is much more easy to handle. As a matter of fact, we have from \eqref{CEXIAA} and
\eqref{DEFRMAA} that ${R_n= n \mu + I_{n-1} \theta}$. Then, \eqref{DECMARTAA} implies that
\begin{equation}
\label{SHARPDECAA}
S_n  = M_n + n \mu + I_{n-1} \theta.
\end{equation}
Finally, if $p<1$ we immediately deduce from \eqref{SHARPDECAA} together with \eqref{CVGMGMN} and \eqref{CVGLN} that
\begin{equation}
\label{CVGSNAA}
\lim_{n \rightarrow \infty} \frac{1}{n}S_n= \mu \hspace{1cm}\text{a.s.}
\end{equation}
More precisely, \eqref{RATECVGMGNN} together with \eqref{RATECVGLN} ensure that
\begin{equation}
\label{RATECVGSNAA}
\Bigl\| \frac{1}{n}S_n - \mu \Bigr\|^2=O\Bigl( \frac{\log n}{n} \Bigr)\hspace{1cm}\text{a.s.}
\end{equation}
which completes the proof of Theorem \ref{T-ASCVGRWGA}.
\demend
%%%%%%%%%%%%%%%%%%%%%%%%%%%%%%%%%%%%%%%%%%%%%%%%%%%%%%%%%%%%%%%%%%%%%%%%%%%%%%%%%%%%%%%%%%%%%%%%%%%%%%%%%%%%%%%%%%%%%%%%
\noindent{\bf Proof of Theorem \ref{T-ANRWGA}.}
%%%%%%%%%%%%%%%%%%%%%%%%%%%%%%%%%%%%%%%%%%%%%%%%%%%%%%%%%%%%%%%%%%%%%%%%%%%%%%%%%%%%%%%%%%%%%%%%%%%%%%%%%%%%%%%%%%%%%%%%
The proof of Theorem \ref{T-ANRWGA} relies on the central limit theorem for multi-dimensional martingales given e.g. 
by Corollary 2.1.10 in (\onlinecite{Duflo97}). 
In the special case where $p=1$, we clearly have $\theta=0$ and $\nu=0$,
which implies from \eqref{IPMN} that $\langle M \rangle_n=n \sigma^2$, and the asymptotic normality trivially holds as
$$
\frac{1}{\sqrt{n}}\bigl( S_n - n \mu \bigr) = \frac{1}{\sqrt{n}} M_n.
$$
Hereafter, we assume that the parameter $0\leq p <1$.
Let $(\cM_n)$ be the martingale with values in $\dR^4$, given by
$$
\cM_n=\begin{pmatrix}
M_n \\
N_n
\end{pmatrix}
$$
where $M_n$ and $N_n$ were previously defined in \eqref{DEFMARTAA} and \eqref{DEFMARTNEWAA}, respectively. Its predictable quadratic variation
$\langle \cM \rangle_n$ can be splited into four terms
$$
\langle \cM \rangle_n=\begin{pmatrix}
\langle M \rangle_n  & \langle C \rangle_n\\
\langle C \rangle_n^T   & \langle N \rangle_n
\end{pmatrix}
$$
where $\langle M\rangle_n$ and $\langle N\rangle_n$ have been previously calculated in \eqref{IPMN} and \eqref{IPNN}, while
\begin{equation*}
\langle C\rangle_n = \sum_{k=1}^n \bigl(\dE[\xi_k \varepsilon_k | \cF_{k-1}]- \dE[\xi_k| \cF_{k-1}]\dE[\varepsilon_k| \cF_{k-1}]\bigr).
\end{equation*}
As before, we have for all $n\geq 0$,
$\dE[ \xi_{n+1}\varepsilon_{n+1} | \cF_n] = \dE[ \xi_{n+1}\varepsilon_{n+1} \vert S_n\in\cV_0]\rI_{S_n\in\cV_0} + 
\dE[ \xi_{n+1} \varepsilon_{n+1}| S_n\in\cV_1]\rI_{S_n\in\cV_1}$.
Hence, we get from \eqref{TransAAV1}, \eqref{TransAAV2} and \eqref{TransAAH} that
\begin{align*}
    \dE [ \xi_{n+1}\varepsilon_{n+1} | \cF_n] =  \sum_{i,j=0}^1 (-1)^i \dE[ \xi_{n+1}\rI_{S_{n+1}\in\cV_i} \vert S_n\in\cV_j]\rI_{S_n\in\cV_j},& \\
    = \sum_{j=0}^1 (-1)^j \Bigg(2 \dE[ \xi_{n+1}\rI_{S_{n+1}\in\cV_j} \vert S_n\in\cV_j]\rI_{S_n\in\cV_j}& \\
    - \dE[ \xi_{n+1} \vert S_n\in\cV_j]\rI_{S_n\in\cV_j}\Bigg).&
\end{align*}
It clearly leads to
\begin{equation}
\label{CEXIVAREPSAA}
 \dE [ \xi_{n+1} \varepsilon_{n+1}| \cF_n]= \bigl(2\zeta -\dE [ \xi_{n+1} | \cF_n]\bigr) \varepsilon_n
\end{equation}
where 
$$
\zeta=\begin{pmatrix}
0\\
0 \\
hp(2\alpha-1)
\end{pmatrix}.
$$
Therefore, we deduce from \eqref{CEXIAA}, \eqref{CEVAREPSAA} and \eqref{CEXIVAREPSAA} that
\begin{align*}
   \langle C \rangle_n  &= \sum_{k=1}^n\bigl(( 2\zeta-\dE[\xi_k\vert\cF_{k-1}]) \varepsilon_{k-1}
   -\dE[\xi_k\vert\cF_{k-1}](2p-1)\varepsilon_{k-1}\bigr), \\
    &= \sum_{k=1}^n \bigl( (2 \zeta-\mu -\theta \varepsilon_{k-1}) \varepsilon_{k-1}-
    (2p-1)( \mu +\theta \varepsilon_{k-1})\varepsilon_{k-1}\bigr), \\
    &= \sum_{k=1}^n \bigl( 2(\zeta -p \mu ) \varepsilon_{k-1}-2p\theta \varepsilon_{k-1}^2\bigr), 
\end{align*}
which implies that
$$
\langle C \rangle_n=-2np\theta+2\bigl(\zeta -p \mu\bigr)I_{n-1}.
$$
Consequently, we immediately obtain from \eqref{CVGLN}
\begin{equation}
\label{CVGCN}
\lim_{n \rightarrow \infty} \frac{1}{n}\langle C \rangle_n= -2p\theta \hspace{1cm}\text{a.s.}
\end{equation}
Hence, it follows from the conjunction of \eqref{CVGIPNN}, \eqref{CVGIPMN} and \eqref{CVGCN} that
\begin{equation}
\label{CVGIPCALMA}
\lim_{n \rightarrow \infty} \frac{1}{n}\langle \cM \rangle_n
=\Lambda=\begin{pmatrix}
\sigma^2 & -2p\theta\\
-2p\theta^T & 4p(1-p) \\
\end{pmatrix}
\hspace{1cm}\text{a.s.}
\end{equation}
In addition, we already saw that the increments of the martingale $(\cM_n)$ are almost surely bounded which ensures that
Lindeberg's condition is satisfied. Hence, we deduce from Corollary 2.1.10 in (\onlinecite{Duflo97}) the asymptotic normality
\begin{equation}
\label{ANCALMN}
\frac{1}{\sqrt{n}}\cM_n \liml \cN \left(0, \Lambda\right).
\end{equation}
Since $p\neq 1$, we get from \eqref{DECMARTNEWAA} that
$I_n = 1+N_n +(2p-1) I_{n-1}$ which implies that $I_{n-1} +\varepsilon_n = 1+N_n +(2p-1) I_{n-1}$, leading to
$$
I_{n-1}=\frac{1+N_n-\varepsilon_n}{2(1-p)}.
$$
Consequently, we have from \eqref{SHARPDECAA} that
\begin{equation}
\label{ANDECAA}
S_n - n \mu = M_n + I_{n-1} \theta=M_n + N_n\theta_p + (1-\varepsilon_n)\theta_p
\end{equation}
where
$$
\theta_p=\frac{1}{2(1-p)}\theta.
$$
The rest of the proof relies on identity \eqref{ANDECAA} together with the well-known Cram\'er-Wold theorem given e.g. by Theorem 29.4 in
(\onlinecite{Billingsley95}). We clearly obtain from \eqref{ANDECAA} that for all $u\in \dR^3$,
\begin{equation}
\label{CWDECAA}
\frac{1}{\sqrt{n}}u^T\bigl( S_n - n \mu \bigr) = \frac{1}{\sqrt{n}} v^T \cM_n  + \frac{(1-\varepsilon_n)}{\sqrt{n}}u^T\theta_p
\end{equation}
where
$$
v=\begin{pmatrix}
u\\
\theta_p^T u \\
\end{pmatrix}.
$$
On the one hand, it follows from \eqref{ANCALMN} that
\begin{equation}
\label{CWAN1AA}
\frac{1}{\sqrt{n}} v^T \cM_n  \liml \cN \left(0, v^T \Lambda v \right).
\end{equation}
On the other hand, as $(1-\varepsilon_n)  \in \{0,2\}$, we immediately have
\begin{equation}
\label{CWAN2AA}
\lim_{n \rightarrow \infty} \frac{(1-\varepsilon_n)}{\sqrt{n}}= 0 \hspace{1cm}\text{a.s.}
\end{equation}
Consequently, we deduce from \eqref{CWDECAA} together with \eqref{CWAN1AA} and \eqref{CWAN2AA} that
\begin{equation}
\label{CWAN3AA}
\frac{1}{\sqrt{n}}u^T\bigl( S_n - n \mu \bigr)  \liml \cN \left(0, v^T \Lambda v \right).
\end{equation}
However, we can easily see from \eqref{CVGIPCALMA} that
$$
v^T\Lambda v =  u^T \sigma^2 u - 4p u^T \theta \theta_p^T u +4p(1-p) u^T \theta_p \theta_p^T u=u^T \Gamma u
$$
where
$$
\Gamma=\sigma^2  -4p  \theta \theta_p^T+4p(1-p) \theta_p \theta_p^T= \sigma^2 - \Bigl( \frac{p}{1-p}\Bigr)\theta \theta^T.
$$
Finally, we find from \eqref{CWAN3AA} and the Cram\'er-Wold theorem that
\begin{equation}
\label{CWAN4AA}
\frac{1}{\sqrt{n}}\bigl( S_n - n \mu \bigr)  \liml \cN \left(0, \Gamma \right)
\end{equation}
which completes the proof of Theorem \ref{T-ANRWGA}.
\demend

%%%%%%%%%%%%%%%%%%%%%%%%%%%%%%%%%%%%%%%%%%%%%%%%%%%%%%%%%%%%%%%%%%%%%%%%%%%%%%%%%%%%%%%%%%%%%%%%%%%%%%%%%%%%%%%%%%%%%%%%
\section{Proofs for the RWG}
\label{S-PRAB}
%%%%%%%%%%%%%%%%%%%%%%%%%%%%%%%%%%%%%%%%%%%%%%%%%%%%%%%%%%%%%%%%%%%%%%%%%%%%%%%%%%%%%%%%%%%%%%%%%%%%%%%%%%%%%%%%%%%%%%%%

%%%%%%%%%%%%%%%%%%%%%%%%%%%%%%%%%%%%%%%%%%%%%%%%%%%%%%%%%%%%%%%%%%%%%%%%%%%%%%%%%%%%%%%%%%%%%%%%%%%%%%%%%%%%%%%%%%%%%%%%
\noindent{\bf Proof of Theorem \ref{T-ASCVGRWGB}.}
%%%%%%%%%%%%%%%%%%%%%%%%%%%%%%%%%%%%%%%%%%%%%%%%%%%%%%%%%%%%%%%%%%%%%%%%%%%%%%%%%%%%%%%%%%%%%%%%%%%%%%%%%%%%%%%%%%%%%%%%
As in the proof of Theorem \ref{T-ASCVGRWGA}, denote by $(\xi_n)$ the increments of the RWG. Then, the position of the RWG 
is given, for all $n \geq 0$, by
\begin{equation}
\label{POSRWG}
S_{n+1}=S_n+\xi_{n+1}
\end{equation}
where
$$
\xi_{n+1}=\begin{pmatrix}
X_{n+1}-X_n \\
Y_{n+1}-Y_n\\
Z_{n+1}-Z_n
\end{pmatrix}.
$$
Let $(\cF_n)$ be the natural filtration associated with the RWG, that is $\cF_n=\sigma\big(\xi_1,\dots,\xi_n\big)$.
We have for all $n\geq 0$,
$$
\dE[ \xi_{n+1} | \cF_n] = \sum_{i=0}^1\sum_{j=0}^1\dE[ \xi_{n+1} \vert S_n\in\cV_{i,j}]\rI_{S_n\in\cV_{i,j}}.
$$
For $i,j=0,1$, denote $u_{i,j}=p_{i,j,1}+p_{i,j,2}$ and $v_{i,j}=p_{i,j,1}-p_{i,j,2}$. Hence, it follows from \eqref{TransABV1}, \eqref{TransABV2} and \eqref{TransABH} that
\begin{align*}
    \dE [ \xi_{n+1} | \cF_n] 
    =& 
    \begin{pmatrix}
    a\bigl( (1-p)-\frac{3}{2}u_{0,0} \bigr) \\
    \frac{a\sqrt{3}}{2} v_{0,0} \\
    hp(2\alpha -1)
    \end{pmatrix}\rI_{S_n\in\cV_{0,0}} \\
    &+ 
    \begin{pmatrix}
    a\bigl( -(1-p)+\frac{3}{2}u_{1,0} \bigr) \\
    -\frac{a\sqrt{3}}{2} v_{1,0} \\
    hp(2\alpha -1)
    \end{pmatrix}\rI_{S_n\in\cV_{1,0}}
    \\
    & +
    \begin{pmatrix}
    a\bigl( 1-\frac{3}{2}u_{0,1} \bigr) \\
    \frac{a\sqrt{3}}{2} v_{0,1} \\
    0
    \end{pmatrix}\rI_{S_n\in\cV_{0,1}} \\
    &+
    \begin{pmatrix}
    a\bigl( -1+\frac{3}{2}u_{1,1} \bigr) \\
    -\frac{a\sqrt{3}}{2} v_{1,1} \\
    0
    \end{pmatrix}\rI_{S_n\in\cV_{1,1}}.
\end{align*} 
This time, it is necessary  to introduce three random variables to discriminate the different vertices. More precisely, let
\begin{align*}
i_n &=\rI_{S_n\in\cV_{0,0}\cup\cV_{0,1}} -\rI_{S_n\in\cV_{1,0}\cup\cV_{1,1}}, \\
j_n &=\rI_{S_n\in\cV_{0,0}\cup\cV_{1,0}} -\rI_{S_n\in\cV_{0,1}\cup\cV_{1,1}}, \\
k_n &=\rI_{S_n\in\cV_{0,0}\cup\cV_{1,1}} -\rI_{S_n\in\cV_{0,1}\cup\cV_{1,0}}.
\end{align*}
The variable $i_n$ keeps track of the local horizontal geometry, while $j_n$ depends on whether or not the particle can jump vertically and $k_n$ 
only depends on the altitude of the particle. Then, $\dE [ \xi_{n+1} | \cF_n]$ reduces to
\begin{align}
    \dE [ \xi_{n+1} | \cF_n] 
    =& 
    \bigl(\mu+\theta+m+\rho\bigr)\rI_{S_n\in\cV_{0,0}} \nonumber \\
    &+ 
    \bigl(\mu-\theta+m-\rho\bigr)\rI_{S_n\in\cV_{1,0}} \nonumber \\
    &  + 
    \bigl(\mu+\theta-m-\rho\bigr)\rI_{S_n\in\cV_{0,1}} \nonumber \\
    &+
    \bigl(\mu-\theta-m+\rho\bigr)\rI_{S_n\in\cV_{1,1}}, \nonumber \\
    =& \mu + \theta i_n + m j_n + \rho k_n
\label{CEXIAB}    
\end{align} 
where the vectors $\mu$, $m$ and $\theta$, $\rho$ are previously defined in \eqref{DEFMUTHETAB} and \eqref{DEFRHOTHETAB}
Moreover, we also have for all $n\geq 0$,
$$\dE[ \xi_{n+1}\xi_{n+1}^T | \cF_n] = \sum_{i=0}^1\sum_{j=0}^1\dE[ \xi_{n+1}\xi_{n+1}^T \vert S_n\in\cV_{i,j}]\rI_{S_n\in\cV_{i,j}}.$$
Hence, we obtain once again from \eqref{TransABV1}, \eqref{TransABV2} and \eqref{TransABH} that
\begin{align*}
    &\dE [ \xi_{n+1}\xi_{n+1}^T | \cF_n] 
    = \\
    &\hspace{0.5cm} \begin{pmatrix}
    a^2\bigl( (1-p) -\frac{3}{4}u_{0,0} \bigr) &  -\frac{a^2 \sqrt{3}}{4}v_{0,0} & 0\\
    -\frac{a^2 \sqrt{3}}{4}v_{0,0} & \frac{3 a^2}{4}u_{0,0} & 0\\
    0 & 0 & h^2 p
    \end{pmatrix}\rI_{S_n\in\cV_{0,0}} 
    \\
    & \hspace{0.5cm}+
    \begin{pmatrix}
    a^2\bigl( (1-p) -\frac{3}{4}u_{1,0} \bigr) &  -\frac{a^2 \sqrt{3}}{4}v_{1,0} & 0\\
    -\frac{a^2 \sqrt{3}}{4}v_{1,0} & \frac{3 a^2}{4}u_{1,0} & 0\\
    0 & 0 & h^2 p
    \end{pmatrix}\rI_{S_n\in\cV_{1,0}} 
    \\
    &  \hspace{0.5cm} + 
    \begin{pmatrix}
    a^2\bigl( 1 -\frac{3}{4}u_{0,1} \bigr) &  -\frac{a^2 \sqrt{3}}{4}v_{0,1} & 0\\
    -\frac{a^2 \sqrt{3}}{4}v_{0,1} & \frac{3 a^2}{4}u_{0,1} & 0\\
    0 & 0 & 0
    \end{pmatrix}\rI_{S_n\in\cV_{0,1}} 
    \\
    & \hspace{0.5cm} +
    \begin{pmatrix}
    a^2\bigl( 1 -\frac{3}{4}u_{1,1} \bigr) &  -\frac{a^2 \sqrt{3}}{4}v_{1,1} & 0\\
    -\frac{a^2 \sqrt{3}}{4}v_{1,1} & \frac{3 a^2}{4}u_{1,1} & 0\\
    0 & 0 & 0
    \end{pmatrix}\rI_{S_n\in\cV_{1,1}}.
\end{align*}
It implies that
\begin{align}
    &\dE[\xi_{n+1}\xi_{n+1}^T | \cF_n] = \nonumber \\
    &\Bigl(\sigma^2+\nu+\gamma+\delta+(\mu+\theta+m+\rho)(\mu+\theta+m+\rho)^T\Bigr)\rI_{S_n\in\cV_{0,0}} \nonumber \\
    & + \Bigl(\sigma^2-\nu+\gamma-\delta+(\mu-\theta+m-\rho)(\mu-\theta+m-\rho)^T\Bigr)\rI_{S_n\in\cV_{1,0}}  \nonumber \\
    & + \Bigl(\sigma^2+\nu-\gamma-\delta+(\mu+\theta-m-\rho)(\mu+\theta-m-\rho)^T\Bigr)\rI_{S_n\in\cV_{0,1}} \nonumber \\
    & + \Bigl(\sigma^2-\nu-\gamma+\delta+(\mu-\theta-m+\rho)(\mu-\theta-m+\rho)^T\Bigr)\rI_{S_n\in\cV_{1,1}},  \nonumber \\
    &= \Bigl(\sigma^2 + \mu\mu^T + \theta\theta^T+ m m^T+ \rho\rho^T\Bigr) \nonumber \\
    &\hspace{0.5cm}+ \Bigl(\nu + \mu\theta^T + \theta\mu^T +m\rho^T +\rho m^T\Bigr) i_n \nonumber \\
    &\hspace{0.5cm}+ 
    \Bigl(\gamma + \mu m^T + m\mu^T +\theta\rho^T +\rho\theta^T\Bigr) j_n \nonumber \\
    &\hspace{0.5cm} + \Bigl(\delta + \mu\rho^T + \rho\mu^T +\theta m^T + m \theta^T\Bigr) k_n  
\label{CEXI2AB}
\end{align}
where the matrices $\sigma^2$ and $\gamma$ are given by \eqref{DEFALPHAGAMMA}, while 
the matrices $\nu$ and $\delta$ are defined by
\begin{equation}
\label{DEFNUDELTA}
\nu = \begin{pmatrix}
\nu_1 & \nu_4 & \nu_6\\
\nu_4 & \nu_2 & \nu_5 \\
\nu_6 & \nu_5 & 0
\end{pmatrix}
\hspace{1cm}\text{and}\hspace{1cm}
\delta = \begin{pmatrix}
\delta_1 & \delta_4 & \delta_6\\
\delta_4 & \delta_2 & \delta_5 \\
\delta_6 & \delta_5 & 0
\end{pmatrix}
\end{equation}
with
\begin{align*}
\nu_1 &=\frac{a^2}{4} \Bigl(\,\frac{9}{4}\bigl(s_{0,0}-s_{1,0}+s_{0,1}-s_{1,1}\bigr)-3p\bigl(u_{0,0}-u_{1,0}\bigr)\Bigr), \\
\nu_2 &=\frac{3a^2}{16}\Bigl( \bigl(u_{0,0}-u_{1,0}+u_{0,1}-u_{1,1}\bigr) - \bigl(v_{0,0}^2-v_{1,0}^2+v_{0,1}^2-v_{1,1}^2\bigr)\Bigr), \\
\nu_4 &=\frac{a^2 \sqrt{3}}{16}\Bigl(3\bigl(t_{0,0}-t_{1,0}+t_{0,1}-t_{1,1}\bigr) +2p\bigl(v_{0,0}-v_{1,0}\bigr)\Bigr), \\
\nu_5 &=\frac{-ah\sqrt{3}}{8}\Bigl(p(2\alpha-1)\bigl(v_{0,0}-v_{1,0}\bigr)\Bigr), \\
\nu_6 &=\frac{ah}{8}\Bigl(3p(2\alpha-1)\bigl(u_{0,0}-u_{1,0}\bigr)\Bigr), 
\end{align*}
and
\begin{align*}
\delta_1 &= \frac{a^2}{4}\Bigl(\,\frac{9}{4}\bigl(s_{0,0}-s_{1,0}-s_{0,1}+s_{1,1}\bigr)-3p\bigl(u_{0,0}-u_{1,0}\bigr)\Bigr), \\
\delta_2 &= \frac{3a^2}{16}\Bigl( \bigl(u_{0,0}-u_{1,0}-u_{0,1}+u_{1,1}\bigr) - \bigl(v_{0,0}^2-v_{1,0}^2-v_{0,1}^2+v_{1,1}^2\bigr)\Bigr), \\
\delta_4 &= \frac{a^2\sqrt{3}}{16}\Bigl(3\bigl(t_{0,0}-t_{1,0}-t_{0,1}+t_{1,1}\bigr) +2p\bigl(v_{0,0}-v_{1,0}\bigr)\Bigr), \\
\delta_5 &= \frac{-ah\sqrt{3}}{8}\Bigl(p(2\alpha-1)\bigl(v_{0,0}-v_{1,0}\bigr)\Bigr), \\
\delta_6 &= \frac{ah}{8}\Bigl(3p(2\alpha-1)\bigl(u_{0,0}-u_{1,0}\bigr)\Bigr), 
\end{align*}
where for all $i=0,1$ and $j=0,1$, $s_{i,j}=u_{i,j}(1-u_{i,j})$ and $t_{i,j}=v_{i,j}(u_{i,j}-1)$.
Therefore, we have the martingale decomposition
\begin{equation}
\label{DECMARTAB}
S_n= \sum_{\ell=1}^n\left(\xi_\ell - \dE [\xi_\ell | \cF_{\ell-1}]\right) + \sum_{\ell=1}^n \dE [\xi_\ell | \cF_{\ell-1}] = M_n+R_n
\end{equation}
where $(M_n)$ is the locally square integrable martingale given by
\begin{equation}
\label{DEFMARTAB}
M_n= \sum_{\ell=1}^n\left(\xi_\ell - \dE [\xi_\ell | \cF_{\ell-1}]\right)
\end{equation}
and the centering term
\begin{equation*}
R_n= \sum_{\ell=1}^n  \dE [\xi_\ell | \cF_{\ell-1}].
\end{equation*}
The predictable quadratic variation associated with
$(M_n)$ is given by
$$
\langle M\rangle_n = \sum_{\ell=1}^n \bigl(\dE[\xi_\ell\xi_\ell^T | \cF_{\ell-1}]-\dE[\xi_\ell| \cF_{\ell-1}]\dE[\xi_\ell| \cF_{\ell-1}]^T\bigr).
$$
We infer from \eqref{CEXIAB} and \eqref{CEXI2AB} that
\begin{align*}
   \langle M\rangle_n &=  \sum_{\ell=1}^n
    \Bigl(\bigl( \sigma^2 + \mu\mu^T + \theta\theta^T+ m m^T+ \rho\rho^T \bigr) \\
    &\hspace{0.9cm}+ 
    \bigl(\nu + \mu\theta^T + \theta\mu^T +m\rho^T +\rho m^T \bigr) i_{\ell-1} \\
    &\hspace{0.9cm}+ 
    \bigl(\gamma + \mu m^T + m\mu^T +\theta\rho^T +\rho\theta^T \bigr) j_{\ell-1} \\
    &\hspace{0.9cm}+ 
    \bigl(\delta + \mu\rho^T + \rho\mu^T +\theta m^T + m \theta^T \bigr) k_{\ell-1} \\
    &\hspace{-1cm}- \bigl(\mu + \theta i_{\ell-1} + m j_{\ell-1} + \rho k_{\ell-1}\bigr)
    \bigl(\mu + \theta i_{\ell-1} + m j_{\ell-1} + \rho k_{\ell-1}\bigr)^T\Bigr).
\end{align*} 
However, it is not hard to see that for all $n \geq 0$, $i_n^2=1$, $j_n^2=1$, $k_n^2=1$ as well as 
$i_nj_n=k_n$, $j_nk_n=i_n$, and $i_nk_n=j_n$.
Consequently, 
$$
\langle M\rangle_n = \sum_{\ell=1}^n \bigl(\sigma^2 + \nu i_{\ell-1} + \gamma j_{\ell-1} + \delta k_{\ell-1} \bigr)
$$
which reduces to
\begin{equation}
\label{NEWIPMN}
\langle M\rangle_n= n \sigma^2  + I_{n-1} \nu  + J_{n-1}\gamma  +  K_{n-1}\delta
\end{equation}
where
$$
I_n=\sum_{\ell=0}^n i_{\ell}, \hspace{1cm} J_n=\sum_{\ell=0}^n j_{\ell}, \hspace{1cm}  K_n=\sum_{\ell=0}^n k_{\ell}.
$$
Furthermore, we clearly have by construction that for all ${n \geq 0}$, $i_n=(-1)^n$ which implies that
\begin{equation}
I_n= \left\lbrace \begin{array}{cc}
1 & \text{if } n \text{ is even},  \vspace{1ex} \\
0 & \text{if } n \text{ is odd}.
\end{array} \right. \label{CEIB}
\end{equation}
In addition, we also have from \eqref{TransABV1}, \eqref{TransABV2} and \eqref{TransABH} that for all $n \geq 0$,
\begin{align} 
   \dE[j _{n+1} | \cF_n] 
    & = (2p-1) \rI_{S_n\in\cV_{0,0}}+(2p-1) \rI_{S_n\in\cV_{1,0}} \\
    &\hspace{0.5cm} +\rI_{S_n\in\cV_{0,1}}+\rI_{S_n\in\cV_{1,1}}, \nonumber \\
    & = 2p \bigl(\rI_{S_n\in\cV_{0,0}}+\rI_{S_n\in\cV_{1,0}}\bigr)-j_n, \nonumber \\
    & = p -(1-p)j_n. 
\label{CEJB}     
\end{align} 
In a similar way, we have for all $n \geq 0$,
\begin{align} 
   \dE[k _{n+1} | \cF_n] 
    & = (1-2p) \rI_{S_n\in\cV_{0,0}}+(2p-1) \rI_{S_n\in\cV_{1,0}}\\
    &\hspace{0.5cm} -\rI_{S_n\in\cV_{0,1}}+\rI_{S_n\in\cV_{1,1}}, \nonumber \\
    & = -2p \bigl(\rI_{S_n\in\cV_{0,0}}-\rI_{S_n\in\cV_{1,0}}\bigr)+k_n, \nonumber \\
    & = (1-p)k_n -pi_n. 
\label{CEKB}     
\end{align}
Consequently, we obtain two more martingale decompositions
\begin{align}
J_n & =  \sum_{\ell=0}^n j_\ell= 1+ \sum_{\ell=1}^n \bigl(j_\ell-\dE[j_\ell | \cF_{\ell-1}] \bigr)+\sum_{\ell=1}^n\dE[j_\ell | \cF_{\ell-1}],
 \nonumber \\
 & = 1+N^J_n + \sum_{\ell=1}^n\bigl( p - (1-p)j_{\ell-1}\Bigr) \nonumber \\
 &=1+N^J_n+ np - (1-p)J_{n-1}, \label{DECMARTNJB}\\
K_n & = \sum_{\ell=0}^n k_\ell= 1+ \sum_{\ell=1}^n \bigl(k_\ell-\dE[k_\ell | \cF_{\ell-1}] \bigr)+\sum_{\ell=1}^n\dE[k_\ell | \cF_{\ell-1}],
\nonumber \\
 & = 1+N^K_n + \sum_{\ell=1}^n \Bigl((1-p)k_{\ell-1} - p i_{\ell-1} \bigr)\nonumber \\
 &=1+N^K_n+ (1-p)K_{n-1}- p I_{n-1},
\label{DECMARTNKB}
\end{align}
where $N^J_n$ and $N^K_n$ are the locally square integrable martingales given by
\begin{align}
\label{DEFMARTNEWABJ}
N^J_n&= \sum_{\ell=1}^n \bigl(j_\ell-\dE[j_\ell | \cF_{\ell-1}] \bigr)\\
\label{DEFMARTNEWABK} N^K_n&= \sum_{\ell=1}^n \bigl(k_\ell-\dE[k_\ell | \cF_{\ell-1}] \bigr).
\end{align}
We already saw that for all $n \geq 0$, $i_n^2=1$, $j_n^2=1$, $k_n^2=1$ and $i_n k_n=j_n$. Hence,
we have from \eqref{CEJB} and \eqref{CEKB} that the predictable quadratic variations associated with
$(N^J_n)$ and $(N^K_n)$ are respectively given by
\begin{align}
\label{IPNJ}
\langle N^J\rangle_n &= 
\sum_{\ell=1}^n \bigl(2p(1-p)(1+j_{\ell-1})\bigr)= 2p(1-p)\bigl(n+J_{n-1}\bigr), \\
\langle N^K\rangle_n &= 
\sum_{\ell=1}^n \bigl(2p(1-p)(1+i_{\ell-1}k_{\ell-1})\bigr)= 2p(1-p)\bigl( n+J_{n-1}\bigr).
\label{IPNK}
\end{align}
It clearly ensures that $\langle N^J\rangle_n$ and $\langle N^K\rangle_n$ are both bounded by $4np(1-p)$ almost surely which immediately implies that
\begin{equation*}
\lim_{n \rightarrow \infty} \frac{1}{n}N^J_n= 0 \hspace{1cm}\text{and}\hspace{1cm}  \lim_{n \rightarrow \infty} \frac{1}{n}N^K_n= 0  \hspace{1cm}\text{a.s.}
\end{equation*}
However, we have from \eqref{DECMARTNJB} and \eqref{DECMARTNKB} that
\begin{eqnarray*}
(2-p)J_n  & = & 1 + N^J_n + np -(1-p)j_n, \\
p K_n & = & 1 + N^K_n -(1-p)k_n - pI_{n-1}.
\end{eqnarray*}
It allows us to deduce that in the case $p>0$,
\begin{align}
\label{CVGIJK}
\lim_{n \rightarrow \infty} \frac{1}{n}I_n= 0, \hspace{0.2cm} \lim_{n \rightarrow \infty} \frac{1}{n}J_n= \frac{p}{2-p}, \hspace{0.2cm}  
\lim_{n \rightarrow \infty} \frac{1}{n}K_n= 0 \hspace{0.2cm} \text{a.s.}
\end{align}
Therefore, it follows from the conjunction \eqref{IPNJ}, \eqref{IPNK} and \eqref{CVGIJK} that
\begin{align}
\label{CVGIPJK}
\lim_{n \rightarrow \infty} \frac{1}{n}\langle N^J\rangle_n= \frac{4p(1-p)}{2-p} , \hspace{.2cm}  
\lim_{n \rightarrow \infty} \frac{1}{n}\langle N^K\rangle_n= \frac{4p(1-p)}{2-p} \hspace{0.2cm} \text{a.s.}
\end{align}
One can also observe that the increments of the martingales $(N_n^J)$ and $(N_n^K)$ are bounded by 2.
Hence, the strong law of large numbers for martingales given in the last part of Theorem 1.3.24 in (\onlinecite{Duflo97}) 
implies that
\begin{equation}
\label{RATENJ}
\bigl(N_n^J\bigr)^2=O(n \log n) \hspace{1cm}\text{a.s.}
\end{equation}
which ensures that
\begin{equation}
\vspace{-1ex}
\label{RATEJ}
\Bigl(\frac{J_n}{n} -\frac{p}{2-p}\Bigr)^2=O\Bigl(\frac{\log n}{n}\Bigr) \hspace{1cm}\text{a.s.}
\end{equation}
In addition, we also have
\begin{equation}
\label{RATENK}
\bigl(N_n^K\bigr)^2=O(n \log n) \hspace{1cm}\text{a.s.}
\end{equation}
Hereafter, we find from \eqref{NEWIPMN} that
\begin{equation}
\label{CVGIPMNB}
\lim_{n \rightarrow \infty} \frac{1}{n}\langle M\rangle_n= \sigma^2 + \Bigl(\frac{p}{2-p}\Bigr) \gamma \hspace{1cm}\text{a.s.}
\end{equation}
Consequently, we obtain from the strong law of large numbers for martingales that
\begin{equation}
\label{CVGMGMNB}
\lim_{n \rightarrow \infty} \frac{1}{n}M_n= 0 \hspace{1cm}\text{a.s.}
\end{equation}
More precisely, by examining each component of the martingale $(M_n)$, it follows from the last part of 
Theorem 1.3.24 in (\onlinecite{Duflo97}) that
\begin{equation}
\label{RATECVGMGNMB}
||M_n||^2=O(n \log n) \hspace{1cm}\text{a.s.}
\end{equation}
As in the proof of Theorem \ref{T-ASCVGRWGA}, $R_n$ is much more easy to handle. 
It follows from \eqref{CEXIAB} that
${R_n=  n\mu  + I_{n-1}\theta  + J_{n-1}m +  K_{n-1}\rho}$.
Then, we infer from \eqref{DECMARTAB} that
\begin{equation}
\label{SHARPDECAB}
S_n  = M_n + n \mu + I_{n-1}\theta  + J_{n-1}m +  K_{n-1}\rho.
\end{equation}
Finally, we immediately deduce from \eqref{SHARPDECAB} together with \eqref{CVGIJK} and \eqref{CVGMGMNB} that
\begin{equation*}
\lim_{n \rightarrow \infty} \frac{1}{n}S_n= \mu + \Bigl(\frac{p}{2-p}\Bigr) m \hspace{1cm}\text{a.s.}
\end{equation*}
More precisely, we find from \eqref{RATEJ} and \eqref{RATECVGMGNMB}  that
\begin{equation}
\label{RATECVGSNAB}
\Bigl\| \frac{1}{n}S_n - \mu - \Bigl(\frac{p}{2-p}\Bigr) m \Bigr\|^2=O\Bigl( \frac{\log n}{n} \Bigr)\hspace{1cm}\text{a.s.}
\end{equation}
which achieves the proof of Theorem \ref{T-ASCVGRWGB} when $p>0$.
In the special case where $p=0$, we have $J_n=I_n$ and $K_n=n$ for every $n\geq 1$. Thus, \eqref{CVGIJK} changes to
\begin{align}
\label{CVGIJK2}
\lim_{n \rightarrow \infty} \frac{1}{n}I_n= 0, \hspace{.2cm} \lim_{n \rightarrow \infty} \frac{1}{n}J_n= 0, \hspace{.2cm}  
\lim_{n \rightarrow \infty} \frac{1}{n}K_n= 1 \hspace{.2cm} \text{a.s.}
\end{align}
which implies that
\begin{equation}
\label{CVGIPMNB2}
\lim_{n \rightarrow \infty} \frac{1}{n}\langle M\rangle_n= \sigma^2 + \delta \hspace{1cm}\text{a.s.}
\end{equation}
Therefore \eqref{CVGMGMNB} and \eqref{RATECVGMGNMB} still hold and we get from \eqref{CVGMGMNB}, \eqref{SHARPDECAB} and \eqref{CVGIJK2}
that
\begin{equation*}
\lim_{n \rightarrow \infty} \frac{1}{n}S_n= \mu + \rho \hspace{1cm}\text{a.s.}
\end{equation*}
\enlargethispage{0.5cm}
More precisely,
\begin{equation*}
\Bigl\| \frac{1}{n}S_n - \mu - \rho \Bigr\|^2=O\Bigl( \frac{\log n}{n} \Bigr)\hspace{1cm}\text{a.s.}
\end{equation*}
which achieves the proof in the special case $p=0$.
\demend

%%%%%%%%%%%%%%%%%%%%%%%%%%%%%%%%%%%%%%%%%%%%%%%%%%%%%%%%%%%%%%%%%%%%%%%%%%%%%%%%%%%%%%%%%%%%%%%%%%%%%%%%%%%%%%%%%%%%%%%%
\noindent{\bf Proof of Theorem \ref{T-ANRWGB}.}
%%%%%%%%%%%%%%%%%%%%%%%%%%%%%%%%%%%%%%%%%%%%%%%%%%%%%%%%%%%%%%%%%%%%%%%%%%%%%%%%%%%%%%%%%%%%%%%%%%%%%%%%%%%%%%%%%%%%%%%%
We shall once again make use of the central limit theorem for multi-dimensional martingales given e.g. 
by Corollary 2.1.10 in (\onlinecite{Duflo97}). In the special case where $p=0$, we obtain with \eqref{CVGIPMNB2}
\begin{equation*}
\frac{1}{\sqrt{n}}M_n  \liml \cN \left(0, \sigma^2+\delta \right)
\end{equation*}
and we immediately get from \eqref{SHARPDECAB} that
\begin{equation}
\label{CWAN4AB2}
\frac{1}{\sqrt{n}}\bigl( S_n - n (\mu + \rho) \bigr)  \liml \cN \left(0, \sigma^2+\delta \right).
\end{equation}
Hereafter, we assume that the parameter $p>0$. Let $(\cM_n)$ be the martingale with values in $\dR^5$, given by
$$
\cM_n=\begin{pmatrix}
M_n \\
N^J_n \\
N^K_n
\end{pmatrix}
$$
where $M_n$, $N^J_n$ and $N^K_n$ were previously defined in \eqref{DEFMARTAB}, \eqref{DEFMARTNEWABJ} and \eqref{DEFMARTNEWABK}. Its predictable quadratic variation
$\langle \cM \rangle_n$ can be splited into nine terms
$$
\langle \cM \rangle_n=\begin{pmatrix}
\langle M \rangle_n  & \langle C \rangle_n &  \langle E \rangle_n\\
 \langle C \rangle_n^T &\langle N^J\rangle_n & \langle D \rangle_n \\
  \langle E \rangle_n^T& \langle D \rangle_n  &\langle N^K\rangle_n 
\end{pmatrix}
$$
where $\langle M\rangle_n$, $\langle N^J\rangle_n$ and $\langle N^K\rangle_n$ have been previously calculated in \eqref{NEWIPMN},
\eqref{IPNJ} and \eqref{IPNK}, while
\begin{eqnarray*}
\langle C\rangle_n &=& \sum_{\ell=1}^n \bigl(\dE[\xi_\ell j_\ell | \cF_{\ell-1}]- \dE[\xi_\ell| \cF_{\ell-1}]\dE[j_\ell| \cF_{\ell-1}]\bigr), \\
\langle D\rangle_n &=& \sum_{\ell=1}^n \bigl(\dE[j_\ell k_\ell | \cF_{\ell-1}]- \dE[j_\ell| \cF_{\ell-1}]\dE[k_\ell| \cF_{\ell-1}]\bigr), \\
\langle E\rangle_n &=& \sum_{\ell=1}^n \bigl(\dE[\xi_\ell k_\ell | \cF_{\ell-1}]- \dE[\xi_\ell| \cF_{\ell-1}]\dE[k_\ell| \cF_{\ell-1}]\bigr).
\end{eqnarray*}
For all $n\geq 0$, we get from \eqref{TransABV1}, \eqref{TransABV2} and \eqref{TransABH} that
\begin{align}
\dE [ \xi_{n+1} j_{n+1} | \cF_n] 
    &=\dE[ \xi_{n+1}\rI_{j_{n+1}=j_n} \vert \cF_n]j_n \nonumber \\
    &\hspace{2.6cm}- \dE[ \xi_{n+1}\rI_{j_{n+1}\neq j_n} \vert \cF_n]j_n, \nonumber \\
    & = 2\dE[ \xi_{n+1}\rI_{j_{n+1}=j_n} \vert \cF_n]j_n - \dE[ \xi_{n+1} \vert \cF_n]j_n,   \nonumber \\
    & = 2 \bigl( \rI_{S_n\in \cV_{0,0}} + \rI_{S_n\in \cV_{1,0}}\bigr) \zeta - \dE[ \xi_{n+1} \vert \cF_n]j_n, \nonumber \\
    & = (1+j_n)\zeta - \dE[ \xi_{n+1} \vert \cF_n]j_n, \label{CEXIJAB}
\end{align}
where $\zeta$ is defined in \eqref{DEFZETA}.
Therefore, we deduce from \eqref{CEXIAB}, \eqref{CEJB} and \eqref{CEXIJAB} that
\begin{align*}
   \langle C \rangle_n  &= \sum_{\ell=1}^n\bigl((1+j_{\ell-1})\zeta - \dE[ \xi_{\ell} \vert \cF_{\ell-1}]j_{\ell-1} \\
   &\hspace{3cm} -\dE[\xi_\ell\vert\cF_{\ell-1}]\bigl(p-(1-p)j_{\ell-1}\bigr)\bigr), \\
    &= \sum_{\ell=1}^n \bigl( (1+j_{\ell-1})\zeta -p(1+j_{\ell-1})\dE[\xi_\ell\vert\cF_{\ell-1}] \bigr), \\
    &\hspace{-0.6cm}= \sum_{\ell=1}^n \bigl( \bigl( \zeta-p(\mu+m)\bigr) \bigl(1+j_{\ell-1}\bigr) -p \bigl(\theta+\rho\bigr) \bigl(i_{\ell-1}+k_{\ell-1}\bigr)
    \bigr), 
\end{align*} 
which implies that
$$
\langle C \rangle_n=\bigl(\zeta -p(\mu+m)\bigr)\bigl(n+J_{n-1}\bigr) -p(\theta+\rho)\bigl(I_{n-1}+K_{n-1}\bigr).
$$
Hence, it follows from \eqref{CVGIJK} that
\begin{equation}
\label{CVGCJN}
\lim_{n \rightarrow \infty} \frac{1}{n}\langle C \rangle_n= \Bigl(\frac{2}{2-p}\Bigr)\bigl( \zeta -p(\mu+m) \bigr)\hspace{1cm}\text{a.s.}
\end{equation}
In a similar way, we infer from \eqref{TransABV1}, \eqref{TransABV2} and \eqref{TransABH} that for all $n\geq 0$
\begin{align}
\dE [ \xi_{n+1} k_{n+1} | \cF_n] 
    &=\dE[ \xi_{n+1}\rI_{k_{n+1}=k_n} \vert \cF_n]k_n - \dE[ \xi_{n+1}\rI_{k_{n+1}\neq k_n} \vert \cF_n]k_n, \nonumber \\
    & = \dE[ \xi_{n+1} \vert \cF_n]k_n - 2\dE[ \xi_{n+1}\rI_{k_{n+1}\neq k_n} \vert \cF_n]k_n ,   \nonumber \\
    & =\dE[ \xi_{n+1} \vert \cF_n]k_n - 2\bigl( \rI_{S_n\in \cV_{0,0}} + \rI_{S_n\in \cV_{1,0}}\bigr)k_n\zeta , \nonumber \\
    & = \dE[ \xi_{n+1} \vert \cF_n]k_n - (i_n+k_n)\zeta. \label{CEXIKAB}
\end{align}
Consequently, we find from \eqref{CEXIAB}, \eqref{CEKB} and \eqref{CEXIKAB} that
\begin{align*}
   \langle E \rangle_n  &= \sum_{\ell=1}^n\bigl(\dE[ \xi_{\ell} \vert \cF_{\ell-1}]k_{\ell-1} - (i_{\ell-1}+k_{\ell-1})\zeta \\
   &\hspace{2.5cm} -\dE[\xi_\ell\vert\cF_{\ell-1}]\bigl((1-p)k_{\ell-1}-pi_{\ell-1}\bigr)\bigr), \\
    &= \sum_{\ell=1}^n \bigl( -\zeta(i_{\ell-1}+k_{\ell-1}) +p(i_{\ell-1}+k_{\ell-1})\dE[\xi_\ell\vert\cF_{\ell-1}] \bigr), \\
    &\hspace{-0.6cm}= \sum_{\ell=1}^n\bigl( \bigl( p(\mu+m)-\zeta\bigr)\bigl(i_{\ell-1}+k_{\ell-1}\bigr) + p(\theta+\rho) \bigl(1+j_{\ell-1}\bigr)\bigr), 
\end{align*}
which leads to
$$
\langle E \rangle_n=\bigl(p(\mu+m)-\zeta \bigr)\bigl(I_{n-1}+K_{n-1}\bigr) +p(\theta+\rho)\bigl(n+J_{n-1}\bigr).
$$ 
Therefore, \eqref{CVGIJK} implies that
\begin{equation}
\label{CVGCKN}
\lim_{n \rightarrow \infty} \frac{1}{n}\langle E \rangle_n= \Bigl(\frac{2p}{2-p}\Bigr)\bigl(\theta + \rho \bigr)\hspace{1cm}\text{a.s.}
\end{equation}
The last term $\langle D \rangle_n$ is much more easy to handle. As a matter of fact, we already saw that for all $n\geq 0$, $j_nk_n=i_n$.
Thus, we deduce from \eqref{CEIB}, \eqref{CEJB}, \eqref{CEKB} together with the elementary fact that $I_{n-1}+I_n=1$ that
\begin{align*}
   \langle D \rangle_n  &= \sum_{\ell=1}^n\Bigl(i_\ell-\bigl(p-(1-p)j_{\ell-1}\bigr)\bigl((1-p)k_{\ell-1}-pi_{\ell-1}\bigr)\Bigr), \\
    &= \sum_{\ell=1}^n \bigl(\bigl(-1+p^2+(1-p)^2\bigr)i_{\ell-1} -2p(1-p)k_{\ell-1} \bigr), \\
    &= \sum_{\ell=1}^n \bigl(-2p(1-p)\bigl(i_{\ell-1} +k_{\ell-1}\bigr) \bigr),
\end{align*}
which means that
$$
\langle D \rangle_n=-2p(1-p)\bigl(I_{n-1}+K_{n-1}\bigr).
$$
Hence, \eqref{CVGIJK} ensures that
\begin{equation}
\label{CVGCJKN}
\lim_{n \rightarrow \infty} \frac{1}{n}\langle D \rangle_n= 0\hspace{1cm}\text{a.s.}
\end{equation}
Consequently, it follows from the conjunction of \eqref{CVGIPJK}, \eqref{CVGIPMNB}, \eqref{CVGCJN}, \eqref{CVGCKN} and \eqref{CVGCJKN} that
\begin{equation}
\label{CVGIPCALMB}
\lim_{n \rightarrow \infty} \frac{1}{n}\langle \cM \rangle_n
=\Lambda
\hspace{1cm}\text{a.s.}
\end{equation}
where the limiting matrix
$$
\Lambda=\frac{1}{2-p}\begin{pmatrix}
(2-p)\sigma^2 + p\gamma & 2\bigl(\zeta-p(\mu+m)\bigr) & 2p(\theta+\rho) \\
2\bigl(\zeta-p(\mu+m)\bigr)^T & 4p(1-p) & 0 \\
2p(\theta+\rho)^T & 0 & 4p(1-p)
\end{pmatrix}.
$$
Moreover, we already saw that the increments of the martingale $(\cM_n)$ are almost surely bounded which ensures that
Lindeberg's condition is satisfied. Whence, we obtain from Corollary 2.1.10 in (\onlinecite{Duflo97}) the asymptotic normality
\begin{equation}
\label{ANCALMNB}
\frac{1}{\sqrt{n}}\cM_n \liml \cN \left(0, \Lambda\right).
\end{equation}
Since $p\neq 0$, we have from \eqref{DECMARTNJB} and \eqref{DECMARTNKB} that
$$
J_{n-1}=\frac{N^J_n+np+1-j_n}{2-p},\hspace{.2cm} K_{n-1}=\frac{N^K_n-pI_{n-1}+1-k_n}{p}.
$$
Consequently, we obtain from \eqref{SHARPDECAB} that
\begin{equation}
\label{ANDECAB}
S_n - n \bigl(\mu+pm_p\bigr) = M_n + N^J_n m_p + N^K_n \rho_p + \cR_n
\end{equation}
where the remainder $\cR_n$ stands for
$
{\cR_n= I_{n-1}\bigl(\theta-\rho\bigr)  + \bigl(1-j_n\bigr)m_p + \bigl(1 -k_n\bigr) \rho_p}
$
with
$$
m_p=\frac{1}{2-p}m \hspace{1cm}\text{and}\hspace{1cm} \rho_p=\frac{1}{p}\rho.
$$
Hereafter, we shall once again make use of the Cram\'er-Wold theorem given e.g. by Theorem 29.4 in
(\onlinecite{Billingsley95}). We have from \eqref{ANDECAB} that for all $u\in \dR^3$,
\begin{equation}
\label{CWDECAB}
\frac{1}{\sqrt{n}}u^T\bigl( S_n - n \bigl(\mu+pm_p\bigr) \bigr) = \frac{1}{\sqrt{n}} v^T \cM_n  + \frac{1}{\sqrt{n}}u^T\cR_n
\end{equation}
where the vector $v\in \dR^5$ is given by
$$
v=\begin{pmatrix}
u\\
m_p^T u \\
\rho_p^T u
\end{pmatrix} 
.
$$
On the one hand, it follows from \eqref{ANCALMNB} that
\begin{equation}
\label{CWAN1AB}
\frac{1}{\sqrt{n}} v^T \cM_n  \liml \cN \left(0, v^T \Lambda v \right).
\end{equation}
On the other hand, as $I_{n}\in \{0,1\}$, $(1-j_n)  \in \{0,2\}$ and ${(1-k_n)  \in \{0,2\}}$, we clearly have for all $u\in\dR^3$
\begin{equation}
\label{CWAN2AB}
\lim_{n \rightarrow \infty} \frac{1}{\sqrt{n}}u^T\cR_n= 0 \hspace{1cm}\text{a.s.}
\end{equation}
Consequently, we obtain from \eqref{CWDECAB} together with \eqref{CWAN1AB} and \eqref{CWAN2AB} that
\begin{equation}
\label{CWAN3AB}
\frac{1}{\sqrt{n}}u^T\bigl( S_n - n (\mu + pm_p) \bigr)  \liml \cN \left(0, v^T \Lambda v \right).
\end{equation}
It is not hard to see from \eqref{CVGIPCALMB} that
\begin{align*}
v^T\Lambda v =  \frac{1}{2-p}\Bigl[ u^T \bigl((2-p)\sigma^2+ p\gamma\bigr) u \hspace{3.5cm} \\
+ 2 u^T \bigl(\zeta-p(\mu+m)\bigr)m_p^T u  
+  2u^T m_p \bigl(\zeta-p(\mu+m)\bigr)^T u\\
 +  2p u^T (\theta+\rho)\rho_p^T u +   2p u^T\rho_p (\theta+\rho)^T u   
 \\
 + 4p(1-p) u^Tm_pm_p^T u + 4p(1-p) u^T\rho_p\rho_p^T u \Bigr],
\end{align*} 
which leads to $v^T\Lambda v=u^T \Gamma u$
where
\begin{align*}
   \Gamma &=\frac{1}{2-p}\Bigl[ (2-p)\sigma^2 + p\gamma + 2\bigl(\zeta-p(\mu+m)\bigr)m_p^T \\
   & \hspace{0.5cm}+ m_p \bigl(\zeta-p(\mu+m)\bigr)^T +  2p (\theta+\rho)\rho_p^T  +   2p \rho_p (\theta+\rho)^T  \\
   & \hspace{0.5cm}  + 4p(1-p) m_p m_p^T + 4p(1-p) \rho_p\rho_p^T \Bigr], \\
   &=\sigma^2+ \frac{p}{2-p}\gamma +
   \frac{2}{(2-p)^2}\Bigl(\bigl(\zeta-p\mu\bigr)m^T + m\bigl(\zeta-p\mu\bigr)^T\Bigr)   \\
   & \hspace{0.5cm} -\frac{4p}{(2-p)^3}m m^T+ \frac{2}{2-p}\bigl(\theta \rho^T + \rho \theta^T\bigr) + \frac{4}{p(2-p)}\rho\rho^T.
\end{align*}
Finally, we deduce from \eqref{CWAN3AB} together with the Cram\'er-Wold theorem that
\begin{equation}
\label{CWAN4AB}
\frac{1}{\sqrt{n}}\bigl( S_n - n (\mu +pm_p) \bigr)  \liml \cN \left(0, \Gamma \right)
\end{equation}
which achieves the proof of Theorem \ref{T-ANRWGB}.
\demend

%%%%%%%%%%%%%%%%%%%%%%%%%%%%%%%%%%%%%%%%%%%%%%%%%%%%%%%%%%%%%%%%%%%%%%%%%%%%%%%%%%%%%%%%%%%%%%%%%%%%%%%%%%%%%%%%%%%%%%%%
\section{Conclusion and perspectives}
\label{S-CP}

This paper explicitly gives a law of large numbers and a central limit theorem for random walks in the three-dimensional hexagonal lattices of ice and graphite.

\medskip

They allow to determine the long time behavior of a particle or a defect site moving on these lattices, provided that the jump probabilities along each direction are explicitly known. This kind of considerations is frequent as this mode of propagation is often used to insert foreign bodies into crystalline structures.

\medskip

These results may be strengthened by determining a speed of convergence using classic results about multidimensional martingales. Future developpements could include the long time behavior of exclusion processes on such lattices, in order to better understand under which conditions several defect sites can coalesce to create a fragility in the structure. The study of the center of mass \cite{LoWade19} of such random walks or elephant random walks \cite{BercuLaulin20} in these lattices may be subjects of interest.

%%%%%%%%%%%%%%%%%%%%%%%%%%%%%%%%%%%%%%%%%%%%%%%%%%%%%%%%%%%%%%%%%%%%%%%%%%%%%%%%%%%%%%%%%%%%%%%%%%%%%%%%%%%%%%%%%%%%%%%%

\section*{Data Availability Statement}

Data sharing is not applicable to this article as no new data were created or analyzed in this study.

%%%%%%%%%%%%%%%%%%%%%%%%%%%%%%%%%%%%%%%%%%%%%%%%%%%%%%%%%%%%%%%%%%%%%%%%%%%%%%%%%%%%%%%%%%%%%%%%%%%%%%%%%%%%%%%%%%%%%%%%
\bibliographystyle{acm}

\end{document}